\documentclass[
reprint,
superscriptaddress,
amsmath,amssymb,
aps,
prx,
floatfix
]{revtex4-2}

\usepackage{graphicx}
\usepackage{dcolumn}
\usepackage{bm}
\usepackage{xcolor}

\begin{document}

\preprint{APS/123-QED}

\title{Extended spin relaxation times of optically addressed telecom defects in silicon carbide}

\author{Jonghoon Ahn}
\affiliation{Materials Science Division, Argonne National Laboratory, Lemont, IL 60439, USA}
\affiliation{Center for Molecular Engineering, Argonne National Laboratory, Lemont, IL 60439, USA}
\author{Christina Wicker}
\affiliation{Pritzker School of Molecular Engineering, University of Chicago, Chicago, IL 60637, USA}
\affiliation{Materials Science Division, Argonne National Laboratory, Lemont, IL 60439, USA}
\author{Nolan Bitner}
\affiliation{Pritzker School of Molecular Engineering, University of Chicago, Chicago, IL 60637, USA}
\affiliation{Materials Science Division, Argonne National Laboratory, Lemont, IL 60439, USA}
\author{Michael T. Solomon}
\affiliation{Pritzker School of Molecular Engineering, University of Chicago, Chicago, IL 60637, USA}
\affiliation{Materials Science Division, Argonne National Laboratory, Lemont, IL 60439, USA}
\affiliation{Center for Molecular Engineering, Argonne National Laboratory, Lemont, IL 60439, USA}
\author{Benedikt Tissot}
\affiliation{Department of Physics, University of Konstanz, D-78457 Konstanz, Germany}
\author{Guido Burkard}
\affiliation{Department of Physics, University of Konstanz, D-78457 Konstanz, Germany}
\author{Alan M. Dibos}
\affiliation{Q-NEXT, Argonne National Laboratory, Lemont, IL 60439, USA}
\affiliation{Center for Molecular Engineering, Argonne National Laboratory, Lemont, IL 60439, USA}
\author{Jiefei Zhang}
\affiliation{Materials Science Division, Argonne National Laboratory, Lemont, IL 60439, USA}
\affiliation{Center for Molecular Engineering, Argonne National Laboratory, Lemont, IL 60439, USA}
\author{F. Joseph Heremans}
\affiliation{Materials Science Division, Argonne National Laboratory, Lemont, IL 60439, USA}
\affiliation{Center for Molecular Engineering, Argonne National Laboratory, Lemont, IL 60439, USA}
\affiliation{Pritzker School of Molecular Engineering, University of Chicago, Chicago, IL 60637, USA}
\author{David D. Awschalom}
    \email[Corresponding author: ]{awsch@uchicago.edu}
\affiliation{Materials Science Division, Argonne National Laboratory, Lemont, IL 60439, USA}
\affiliation{Center for Molecular Engineering, Argonne National Laboratory, Lemont, IL 60439, USA}
\affiliation{Pritzker School of Molecular Engineering, University of Chicago, Chicago, IL 60637, USA}
\affiliation{Department of Physics, University of Chicago, Chicago, IL 60637, USA}

\date{\today}

\begin{abstract}

Optically interfaced solid-state defects are promising candidates for quantum communication technologies. The ideal defect system would feature bright telecom emission, long-lived spin states, and a scalable material platform, simultaneously. Here, we employ one such system, vanadium ($V^{4+}$) in silicon carbide (SiC), to establish a potential telecom spin-photon interface within a mature semiconductor host. This demonstration of efficient optical spin polarization and readout facilitates all-optical measurements of temperature-dependent spin relaxation times ($T_1$). With this technique, we lower the temperature from about 2K to 100 mK to observe a remarkable four-orders-of-magnitude increase in spin $T_1$ from all measured sites, with site-specific values ranging from 57 ms to above 27 s. Furthermore, we identify the underlying relaxation mechanisms, which involve a two-phonon Orbach process, indicating the opportunity for strain-tuning to enable qubit operation at higher temperatures. These results position $V^{4+}$ in SiC as a prime candidate for scalable quantum nodes in future quantum networks.

\end{abstract}

\maketitle


\section{\label{sec:intro}Introduction}

A spin-photon interface providing the connection between stationary and flying qubits is an essential requirement for large-scale quantum networks. Optically addressable solid-state spin defects have emerged as a promising contender for such a quantum interface, combining long-lived memory spin qubits with spin-dependent optical transitions to enable long-distance spin entanglement \cite{wolfowicz2021quantum,hermans2022qubit,knaut2023entanglement}. The ideal spin impurity would exhibit bright emission in the telecom band along with favorable spin properties, but few systems possess these qualities together. While sophisticated engineering approaches, such as enhancement of the photon emission with optical cavities and frequency downconversion via nonlinear optical processes \cite{dibos2018atomic,ourari2023indistinguishable,kurkjian2021optical,dreau2018quantum,bersin2024telecom}, have been explored, there remains significant motivation for the search of novel spin defects that inherently satisfy these requirements.

Among various solid-state telecom emitters, vanadium ($V^{4+}$) in silicon carbide (SiC) has recently garnered significant interest as a potential qubit system for quantum technologies. Notable attributes include the short optical lifetimes (11 - 167 ns), emission in the telecom O-band (1278 – 1388 nm), spin-dependent optical transitions, and encouraging spin properties at cryogenic temperatures \cite{spindlberger2019optical,wolfowicz2020vanadium,cilibrizzi2023ultra,astner2022vanadium,hendriks2022coherent}. Furthermore, $V^{4+}$ in SiC is readily available as a conventional compensation dopant, and the mature electronic material platform of SiC is seamlessly compatible with scalable device integration \cite{lukin20204h,anderson2019electrical,whiteley2019spin}.

In this work, we utilize unprocessed commercial SiC wafers with $V^{4+}$ to establish key prerequisites of a telecom spin-photon interface: optical spin polarization (initialization) and readout. We first examine the charge properties under various optical illumination conditions to determine the experimental parameters that exhibit minimal ionization effects. Upon ensuring the charge stability, we harness the intrinsic short optical lifetimes to obtain efficient spin polarization and use the spin-dependent photoluminescence for optical readout of the spin sublevels. Each of these experiments span mulitiple distinct sites of $V^{4+}$ in both 4H- and 6H-SiC, offering versatility to select the optimal site for future quantum communication applications.

As the spin relaxation time, or spin $T_1$, sets a fundamental limit for the spin coherence time ($T_2$), a comprehensive understanding of the underlying physics governing the relaxation dynamics is crucial. Leveraging the optical spin polarization and readout capabilities, we employ all-optical hole burning recovery measurements to systematically investigate the spin $T_1$ of $V^{4+}$ in SiC. By decreasing the sample temperature from approximately 2 K to 100 mK, we discover a remarkable increase in spin $T_1$ spanning nearly four orders of magnitude across all measured sites, yielding site-dependent values ranging from 57.1 ms to 27.9 s. This shows that the spin lifetimes are not prohibitive for coherence at sub-Kelvin temperatures and supports the viability of $V^{4+}$ as a telecom spin qubit. Moreover, the pronounced temperature dependence of the spin $T_1$ reveals the phonon-mediated relaxation mechanisms. Specifically at above 1 K, an Orbach process coupled to the orbital upper-branch ground state is identified as the dominant source for relaxation. Based on these insights, we propose a strain engineering approach to manipulate the orbital levels and improve spin $T_1$, thereby extending the operational temperature range for $V^{4+}$ spins in SiC to a more accessible domain compatible with standard cryogenic systems. Overall, these findings underscore the significant potential of $V^{4+}$ in SiC, providing a promising avenue towards the foundational elements of future quantum networks using commonly available electronic materials.

\section{\label{sec:optical}Optical spectroscopy of $V^{4+}$ in SiC}

\begin{figure}[hbtp]
\centering
\includegraphics[width=1\linewidth]{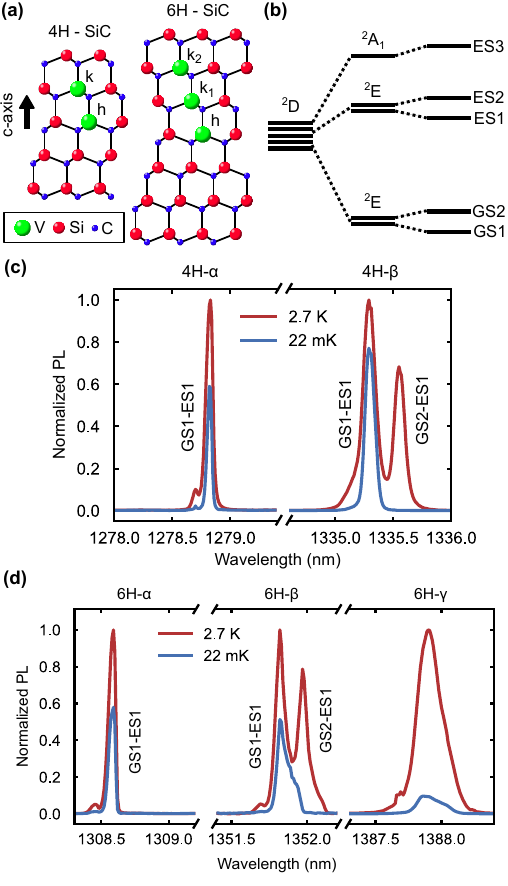}
\caption{\label{fig:PLE} Optical spectroscopy of $V^{4+}$ in 4H- and 6H-SiC. (a) Vanadium defects substitute inequivalent Si sites in the 4H- and 6H-SiC lattice. (b) Orbital level structure of $V^{4+}$ in SiC. The crystal field from the SiC lattice splits the orbital $d^1$ states to an orbital doublet ground state and an excited state consisting of an orbital singlet and doublet state. The order of the orbital singlet and doublet depends on the site. Additionally, the spin orbit interactions lead to SO splitting in the orbital doublets. (c, d) Resonant PLE spectroscopy measurements at 2.7 K (red) and 22 mK (blue) for 4H- and 6H-SiC. The optical transitions in the 6H-$\gamma$ site are within the inhomogeneous linewidth, and thus selective excitation on the GS1-ES1 transition is challenging.}
\end{figure}

Vanadium substitutes for a silicon (Si) atom in the SiC lattice, and thus occupies two inequivalent sites for the 4H polytype (one hexagonal ($h$) and one quasi-cubic ($k$) site) and three inequivalent sites for the 6H polytype (one hexagonal ($h$) and two quasi-cubic ($k_1$, $k_2$) sites) (Fig. \ref{fig:PLE} (a)). The neutral charge state of vanadium ($V^{4+}$) possesses one active electron (spin $1/2$) in its $3d^1$ orbital, forming a free ion ${}^2D$ state. Due to the crystal field of the SiC lattice, the orbital levels split into an orbital doublet ground state and a triplet excited state, with the latter further separated into an orbital singlet and doublet due to the $C_{3v}$ symmetry in 4H- and 6H-SiC \cite{kunzer1993magnetic,kaufmann1997crystal,baur1997transition}. The spin-orbit (SO) interactions transform the orbital singlets into a Kramers doublet (KD) and each doublet into a pair of KDs, separated by a SO splitting \cite{tissot2021spin,tissot2022nuclear}. Depending on the $V^{4+}$ site in each polytype, the site symmetry affects the order of the orbital levels as well as the magnitude of the SO splittings \cite{spindlberger2019optical,csore2020ab}. Throughout this work, in accordance with previous studies \cite{wolfowicz2020vanadium,csore2020ab}, we refer to each site as 4H-$\alpha$ and $\beta$, and 6H-$\alpha$, $\beta$, and $\gamma$ and denote each orbital level in the ground and excited states as GS1 and GS2 and ES1, ES2, and ES3 in ascending order (Fig. \ref{fig:PLE} (b)).

First, we experimentally investigate the orbital structure of the various sites of $V^{4+}$ with resonant photoluminescence excitation (PLE) spectroscopy in the dilution refrigerator at 2.7 K (for setup details, see the Supplemental Material \cite{supp}). As in previous experiments \cite{wolfowicz2020vanadium}, we use a tunable laser for resonant excitation and collect photoluminescence (PL) from the phonon sideband while filtering out the resonant laser. The PLE spectra displays clearly resolved optical transitions (Fig. \ref{fig:PLE} (c), (d)), allowing us to confirm the SO splitting of each site (see Table \ref{tab:Table} and the Supplemental Material \cite{supp}), related to the phonon-relaxation mechanism as we discuss in the following sections (`Phonon mediated spin relaxation' section). Furthermore, the site assignment and the irreducible representation of each orbital level can be determined from the orbital structure and the selection rules (see the Supplemental Material \cite{supp}) except for the 6H-$\gamma$ site where the small SO splitting comparable to its inhomogeneous linewidth makes it a challenge to optically resolve the distinct orbital levels (Fig. \ref{fig:PLE} (d)).

We next lower the sample temperatures to 22 mK and observe a strong thermal polarization in the 4H-$\beta$ and 6H-$\beta$ sites that suppresses the allowed GS2-ES1 transition below the noise level. Similar orbital polarization is already obtained at 2.7 K 4H-$\alpha$ and 6H-$\alpha$ sites due to their larger GS1-GS2 splittings. Additionally, Fig. \ref{fig:PLE} (c) and (d) show that the PLE amplitude for the GS1-ES1 transitions in all sites decrease as the temperature is reduced to below 3 K (see the Supplemental Material \cite{supp}). This reduction could result from temperature-dependent effects altering the emission in the phonon sideband or an increase in the electron spin lifetime \cite{green2017neutral,bosma2018identification,bergeron2020silicon}. The PLE measurements are conducted at zero magnetic field, but a finite energy splitting exists due to hyperfine interactions, and the resonant excitation may pump a sub-population of defects into a long-lived dark state no longer resonant with the excitation frequency. Hence, the reduction in PL emission may provide an initial indication of long lifetimes and narrow linewidths at the sub-Kelvin temperatures. The 6H-$\gamma$ site displays an asymmetric PLE spectrum, indicating that multiple ES states are positioned within the bandgap, while identifying the full orbital structure for this site remains challenging. Due to the difficulty in selective excitation of the optical transitions, we exclude the 6H-$\gamma$ site from the remainder of the study where we rely on the GS1-ES1 transition for resonant excitation of the $\alpha$ and $\beta$ sites for both 4H- and 6H-SiC.

\section{\label{sec:charge}Charge Properties}

\begin{figure*}[!hbtp]
\centering
\includegraphics[width=1\linewidth]{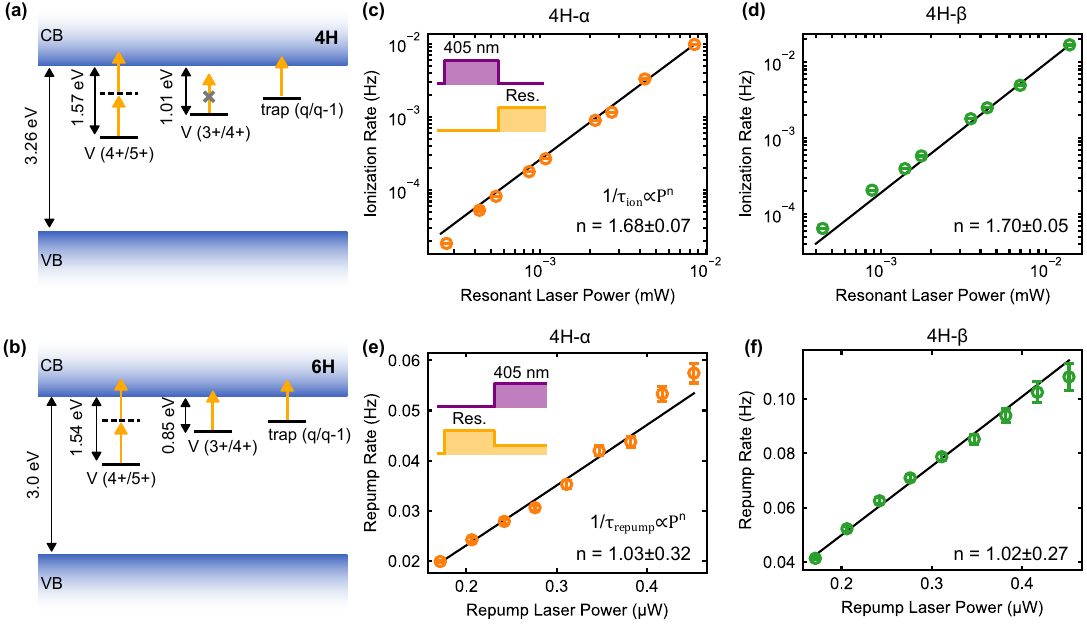}
\caption{\label{fig:Charge} Charge properties of $V^{4+}$ in SiC. (a, b) Acceptor and donor levels of vanadium in 4H- and 6H-SiC reproduced from \cite{mitchel2007vanadium}. Orange arrows illustrate the photon energies from GS1-ES1 resonant excitation. VB and CB correspond to the valence band and conduction band, respectively. (c, d) Resonant laser power dependent ionization rates for 4H-$\alpha$ and 4H-$\beta$ sites. For both sites, the ionization rates are below $10^{-4}$ Hz for laser powers below 500 nW, and the power law results in an exponent n of $\sim$1.7. Inset shows pulse sequence for corresponding measurement. (e, f) Repump rate as a function of 405 nm laser power for 4H-$\alpha$ and 4H-$\beta$ sites. The repump rate is approximately 0.05 and 0.1 Hz at 400 nW and therefore repump times of about 100 s is sufficient for charge repump into the $V^{4+}$ state. The power dependence follows a linear exponent of n $\sim$1 for both sites. Inset shows pulse sequence for corresponding measurement. All reported errors represent 1 SE from the fit, and all error bars represent 1 SD from experimental acquisition.}
\end{figure*}

The charge stability serves as a critical factor that affects the spin state, and thus, in this section we examine the charge properties of $V^{4+}$ ensemble under various laser illumination conditions. With the amphoteric properties of vanadium impurities in SiC, both the $V^{3+}$/$V^{4+}$ acceptor and $V^{4+}$/$V^{5+}$ donor levels reside within the SiC bandgap (Fig. \ref{fig:Charge} (a), (b) reproduced from \cite{mitchel2007vanadium}) and the vanadium atom can be stable in one of three charge states: $V^{3+}$, $V^{4+}$, or $V^{5+}$. While the $V^{3+}$ charge state could be of interest in future studies as a spin-1 system with optical transitions near 2 $\mu$m \cite{von2019transition}, the primary focus of this study is the $V^{4+}$ charge state. Therefore, we identify experimental conditions that minimize the charge effects during the all-optical spin measurements in subsequent sections. It should be noted that while the charge stability of low dose $V^{4+}$ has been recently investigated, the behavior can be drastically different from the $V^{4+}$ ensembles due to the different dopant concentrations within the sample \cite{cilibrizzi2023ultra}.

Similar to previous experiments with $V^{4+}$ ensembles in 4H-SiC, we observe a slow quenching of PL under a continuous resonant laser illumination, a phenomenon absent in 6H-SiC, as well as a PL recovery when a 405 nm repump laser is applied to the sample \cite{wolfowicz2020vanadium,cilibrizzi2023ultra,astner2022vanadium,wolfowicz2017optical}. Our detection scheme is only sensitive to the emission from the $V^{4+}$ charge state, and therfore the reduction in PL will occur as the resonant excitation converts the $V^{4+}$ to one of the unobserved ``dark" charge states ($V^{3+}$ or $V^{5+}$) while the PL recovery indicates a restoring of the population back into the ``bright" $V^{4+}$ charge state. First, we characterize the resonant laser-induced photoionization rates as a function of laser powers. The experimental sequence consists of 405 nm laser exposure of $>$100 s to restore the $V^{4+}$ charge state and a subsequent step where the 405 nm laser is turned off and the resonant laser with different powers are applied to the 4H-$\alpha$ and $\beta$ sites. The photoionization process is examined through the time-dependent decay of PL signal, from which an ionization rate of below $10^{-4}$ Hz is extracted at powers below 500 nW (Fig. \ref{fig:Charge} (c), (d)) \cite{wolfowicz2017optical,magnusson2018excitation}.

Additional studies would be required to fully resolve the mechanisms for the observed photoionization, but there are two potential explanations for the process (Fig. \ref{fig:Charge} (a), (b)). One possible candidate is a two-photon ionization process wherein electrons from the $V^{4+}$/$V^{5+}$ donor state is excited into the conduction band converting a $V^{4+}$ into a $V^{5+}$ charge state. The other possibility is the presence of a donor trap state located close to the conduction band, where the resonant laser energy is sufficient to induce photoionization and create free electrons that are subsequently recaptured by the $V^{4+}$, forming the $V^{3+}$ charge state. According to Secondary Ion Mass Spectrometry (SIMS) analysis (see Materials and Methods for details), our samples have larger concentration of nitrogen (N) compared to boron (B) and aluminum (Al) dopants, which may provide the shallow donor states. The dependence of the ionization rate on the laser power shows a relation of $1/\tau_{ionization} \propto P^{n}$, where the exponent $n$ is $\sim$1.7 for both sites. While a $n$ value close to 2 would provide direct evidence for a two-photon ionization at our low laser intensities, our results remain inconclusive \cite{magnusson2018excitation,anderson2022five}. On the other hand, in either case, the absence of photoionization in the 6H polytype can be explained. According to Ref. \cite{mitchel2007vanadium} (Fig. \ref{fig:Charge} (a), (b), the donor and acceptor levels are respectively at $E_c-1.57$ eV and $E_c-1.01$ eV for the 4H-SiC, and at $E_c-1.54$ eV and $E_c-0.85$ eV for the 6H-SiC where $E_c$ is the conduction band energy. This shows that the photon energies from the resonant excitation (0.89 - 0.97 eV) is sufficient to convert $V^{3+}$ to $V^{4+}$ in the 6H polytype. Therefore, even if the same ionization processes exist in 6H-SiC, the resonant excitation will have sufficient photon energy for the faster one-photon conversion of $V^{3+}$ to $V^{4+}$, resulting in the absence of any significant reduction of PL due to photoionization under resonant excitation.

We next characterize the charge repump rates with the 405 nm laser illumination. The experiments are performed with a sequence consisting of a long ($>$100 s), $\sim$100 $\mu$W resonant laser pulse to ionize the ensemble into the dark state, followed by a reduction in resonant laser power to $\sim$75 nW (needed for PL readout) while the 405 nm laser with different powers are applied to the sample. Repump rates of about 20 to 100 mHz are extracted from the PL recovery traces, and the repump rates show a near linear dependence to the repump laser power ($1/\tau_{repump} \propto P^{n}, n\sim1$) (Fig.\ref{fig:Charge} (e), (f)). This result can be interpreted as a direct photoionization of the dark charge state into the $V^{4+}$ state. Lastly, we characterize the ionization processes where the sample is not under any optical illumination, by applying a 405 nm laser pulse followed by a resonant laser pulse separated with a variable dark time. Results show negligible reduction in PL for periods of up to 10,000 seconds, indicating that the charge state is stable in the dark (see the Supplemental Material \cite{supp}). Overall, the measurements above suggest that all-optical spin measurements are unlikely to have significant charge implications when the experimental excitation parameters are carefully selected. Furthermore, with proper tuning of the doping conditions, it should be possible to achieve a charge stability that does not hinder the potential applications of $V^{4+}$ in SiC.

\section{\label{sec:polarizationreadout}Optical spin polarization and readout}

\begin{figure}[hbtp]
\centering
\includegraphics[width=1\linewidth]{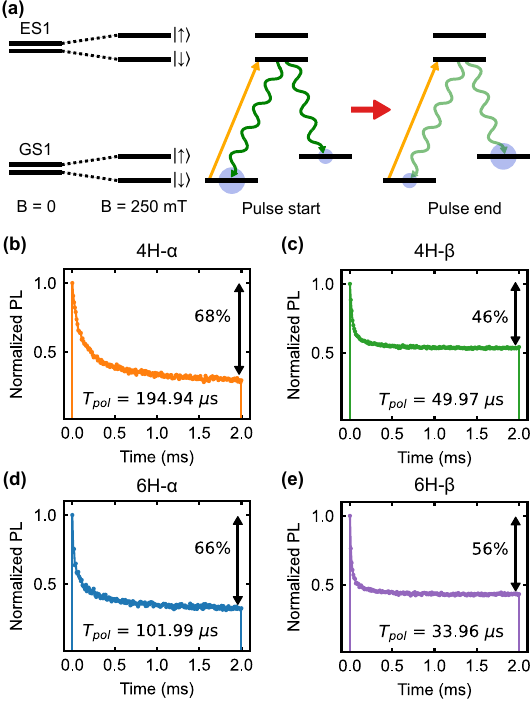}
\caption{\label{fig:Polarization} Optical spin polarization. (a) Under an external magnetic field and at about 23 mK, the Zeeman splittings enables a selective drive of a spin sublevel and allows optical spin polarization. Following the resonant excitation, a subpopulation is pumped into an inaccessible spin sublevel resulting in reduction of PL and optical spin polarization. (b-e) Optical spin polarization is achieved within 2 ms of resonant laser excitation on GS1-ES1 transition. Inset includes polarization timescale and optical contrast achieved in current configuration. Laser pulse length is determined to be sufficiently longer than polarization timescale to allow sufficient population accumulation in the dark state. The PL contrast provides a lower bound on spin initialization.}
\end{figure}

Having demonstrated that the charge state of the $V^{4+}$ ensemble can be maintained stable under specific laser illumination conditions, we proceed to implement optical spin polarization and readout, which are essential requirements for a potential spin-photon interface. Previous studies have shown that $V^{4+}$ in SiC exhibits short optical lifetimes that allow access to single defects without requiring Purcell enhancement, a capability not achievable for many other telecom dopants/defects \cite{dibos2018atomic,ourari2023indistinguishable,kurkjian2021optical}. This short optical lifetime is also advantageous for optical spin polarization. When the optical lifetime is substantially shorter than the spin relaxation time, the system can undergo multiple optical cycles where the spin nonconserving transitions result in a spin flip, and population can be accumulated into the other spin sublevel. Moreover, if the other spin sublevel is no longer resonant with the optical drive, PL is reduced as population moves into the dark state, allowing optical readout of the spin (Fig. \ref{fig:Polarization} (a)). At the sub-Kelvin temperatures used in this study, the spin relaxation times of $V^{4+}$ in SiC easily surpasses the optical lifetimes of 11 to 167 ns, enabling an efficient spin polarization and readout.

We perform optical spin polarization and readout under an external magnetic field of 250 mT applied parallel to the c-axis of the SiC at about 23 mK. This induces a Zeeman splitting of $g\mu_{B} B$ for both the ground and excited state, where $g$ is the Kramers doublet g-factor, $\mu_{B}$ is the Bohr magneton, and $B$ is the static magnetic field. The g-factors differ between the ground and excited state \cite{wolfowicz2020vanadium}, and as the difference in the Zeeman splittings exceeds the subensemble linewidth under the applied 250 mT magnetic field, the resonant laser selectively drives a spin sublevel within the broadened ensemble. This optical pumping process initiates population transfer into the other dark spin sublevel, also referred to as hole burning. Under these conditions, we apply a 2 ms resonant laser pulse to each $V^{4+}$ site, which is short enough that the photoionization effects are negligible. We observe a gradual decrease in PL emission during the applied laser pulse, obtaining spin polarization and readout, as depicted in Fig. \ref{fig:Polarization} (b)-(e).

The transient PL signal acquired above is fit to a single exponential decay, from which we define the polarization timescale, $T_{pol}$. Depending on the site, $T_{pol}$ ranges from 34 to 195 $\mu$s, indicating that the duration of the resonant laser pulse is sufficiently long to accumulate the spin polarization. Furthermore, by comparing $T_{pol}$ to the optical lifetime ($T_{opt}$), we can estimate the strength of the spin non-conserving transition that enables the optical spin polarization. While the orbital transitions are spin conserving to the leading order, the spin-orbit coupling and the hyperfine interaction induce a weak mixing of the spin states, allowing for spin non-conserving transitions \cite{tissot2021spin,tissot2021hyperfine} (details provided in the Supplemental Material \cite{supp}). We perturbatively calculate the strength of each process and find that the hyperfine mixing is the dominant interaction for all measured sites, notably with the estimated mixing being of the same order of magnitude as $T_{opt}/T_{pol}$ (see Table S1). This leads us to conclude that the spin polarization primarily arises from hyperfine mixing of the spin states, which has also been shown to facilitate the otherwise forbidden microwave driving of the spin states \cite{gilardoni2021hyperfine}. Moreover, the optical contrast between the beginning and end of the laser pulse provides a lower bound on the spin polarization, indicating a spin polarization of at least 44 to 68\% has been achieved in our current configuration \cite{bayliss2020optically, diler2020coherent}. The initialization could be improved in future experiments by either using a larger magnetic field or a sample with narrower sub-ensemble linewidth (i.e., isotopically purified SiC or lower $V^{4+}$ density samples). However, the acquired spin polarization is already sufficient for hole burning recovery measurements in the subsequent section.

\section{\label{sec:relaxation}Spin relaxation times at sub-Kelvin temperatures}

With the efficient optical spin polarization and readout, we can now all-optically characterize the spin $T_1$ at sub-Kelvin temperatures. In the solid state, the spin $T_1$ is typically governed by the spin-lattice relaxation, which can ultimately limit the operation temperature of the spin qubit. Therefore investigating the temperature-dependent $T_1$ is crucial for assessing the applicability of $V^{4+}$ in SiC. Recent studies have shown that the spin $T_1$ of $V^{4+}$ in the 4H-$\alpha$ site can be extended to approximately 25 s at $\sim$100 mK \cite{astner2022vanadium}. Here, we employ hole-burning recovery measurements to systematically study the spin $T_1$ in the $\alpha$ and $\beta$ sites for both 4H- and 6H-SiC at sub-Kelvin temperatures.

\begin{figure*}[!hbtp]
\centering
\includegraphics[width=1\linewidth]{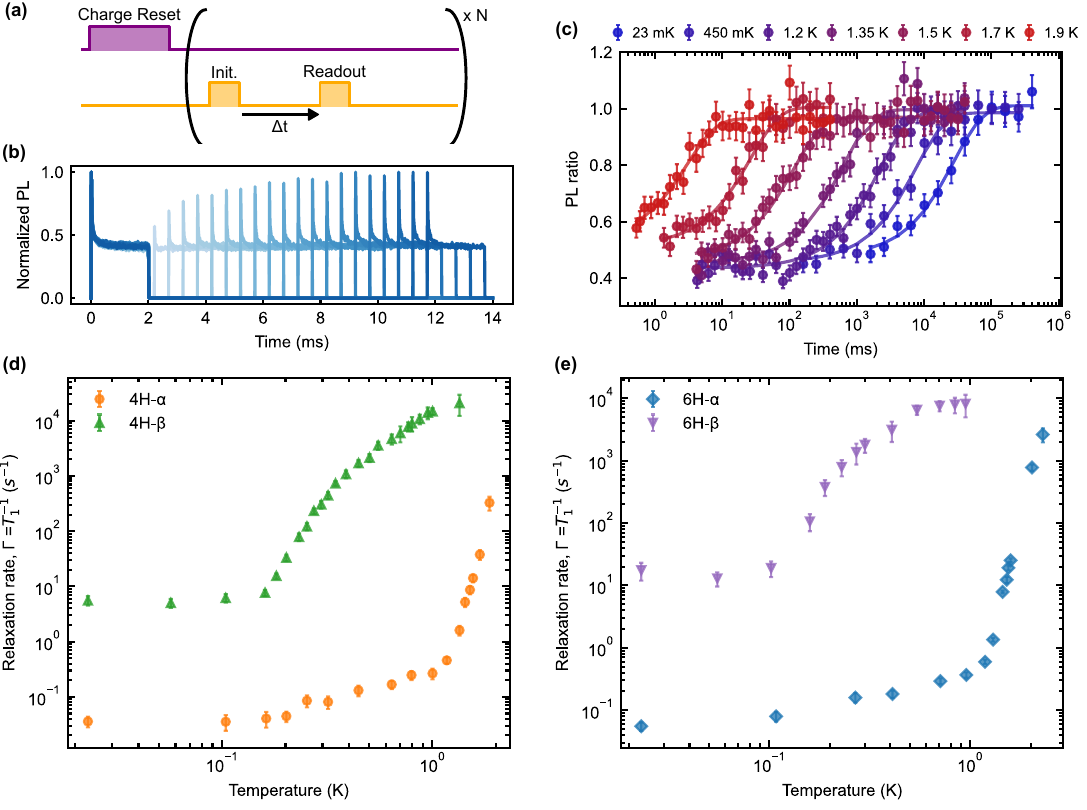}
\caption{\label{fig:Relaxation} Hole burning recovery measurements and temperature dependent spin relaxation times. (a) Pulse sequence of the hole burning recovery experiments. Charge reset laser is required for the 4H polytype to ensure repump into the $V^{4+}$ charge state. Two resonant laser pulses with variable delay time in between are applied for spin polarization and readout of the recovery during delay. After the second resonant laser pulse, a wait time longer that 5 times the spin $T_{1}$ is given to obtain thermal equilibrium before applying subsequent pulse sequence. (b) Example of a hole burning recovery measurement. Readout pulse shows a single exponential PL increase with variable delay time. Time constant of the exponential recovery corresponds to the spin $T_{1}$. (c) Example of temperature dependent spin relaxation measurements of the 4H-$\alpha$ site. Measurements show a spin $T_{1}$ ranging approximately four orders of magnitude between temperatures of 22 mK and 1.9 K. (d, e) Measured temperature dependent spin relaxation rates ($T_{1}^{-1}$) of $\alpha$ and $\beta$ sites in 4H- and 6H-SiC. All reported errors represent 1 SE from the fit, and all error bars represent 1 SD from experimental acquisition.}
\end{figure*}

Fig. \ref{fig:Relaxation} (a) shows the experimental sequence for the hole-burning recovery measurements we use to characterize the longitudinal spin relaxation times. Before each measurement, for 4H-SiC, the $V^{4+}$ charge state is recovered with a 405 nm laser pulse while this step is not necessary and omitted for 6H-SiC. Once we ensure the $V^{4+}$ charge state, a 2 ms resonant laser pulse is applied to induce a spin polarization, or initialization, followed with a variable delay time and a readout pulse. Comparing the PL from the readout to that of the initialization, we probe the amount of population that relaxes back into the “bright” state during the variable delay. The resulting PL displays an exponential increase as a function of the variable delay time and the single time constant is consistent with the spin $T_1$ (Fig. \ref{fig:Relaxation} (b). After each readout pulse, a sufficient wait time ($>$ 5 times longer than spin relaxation time) is given so that the spin state fully recovers to its thermal equilibrium before the next pulse sequence is applied \cite{gilardoni2020spin}. During the experiments, a 250 mT magnetic field is applied along the SiC c-axis to enable the spin polarization.

Next, we repeat the experiments while adjusting the sample temperatures and investigate the temperature dependence of the spin $T_1$. It is worth noting that in the temperature range of our experiments, even the shortest measured $T_1$ is on the order of 10 $\mu$s which is significantly longer than the optical lifetimes and an efficient spin polarization is achieved where the spin $T_1$ exceeds the optical lifetime. Results show a profound $T_1$ dependence on the lattice temperature; for example, the 4H-$\alpha$ site displays a spin $T_1$ increasing from 3.1 ms to 27.9 s when lowering the sample temperatures from 1.9 K to 23 mK, spanning almost four orders of magnitude (Fig. \ref{fig:Relaxation} (c)). The measured spin $T_1$ as a function of temperature are displayed in Fig. \ref{fig:Relaxation} (d) and (e) as relaxation rates ($\Gamma = T_1^{-1}$). These results show that a similar strong temperature dependence is present for all measured sites: both $\alpha$ and $\beta$ sites in 4H- and 6H-SiC. At the base temperature of 23 mK, the 4H-$\alpha$ and $\beta$ sites each show a spin $T_1$ of 27.9 s and 179.5 ms, respectively while the 6H-$\alpha$ and $\beta$ have spin $T_1$ of 18.2 s and 57.1 ms, respectively. Moreover, for the 4H-$\beta$ site, it is also found that the $T_1$ can be extended to above 1 s by reducing the magnitude of the external magnetic field (see the Supplemental Material \cite{supp}). While this effect should also be reproducible for the other sites, it should be emphasized that the current reported values of $T_1$ are sufficient to not restrict the coherence time of $V^{4+}$ in SiC at sub-Kelvin temperatures for applications in large-scale quantum networks.

Our findings clearly prove that the interactions between the electron and the $\sim$100 \% abundant central nuclear spin-7/2 of $V^{4+}$ in SiC are not prohibitive for $T_1$. The dipolar coupling is expected to enhance the spin flips from the interaction with the lattice phonons, but we demonstrate a prolonged $T_1$ is achieved at low temperatures while an efficient electron spin control via microwave driving should still be enabled with the hyperfine-mediated transitions \cite{gilardoni2020spin,gilardoni2021hyperfine,wolfowicz2020vanadium}. Consequently, $V^{4+}$ in SiC satisfies the fundamental criteria as a prospective telecom spin qubit, offering the ability for coherent microwave spin control alongside optical initialization and readout.

\section{\label{sec:phonon}Phonon mediated spin relaxation}

\begin{figure*}[!hbtp]
\centering
\includegraphics[width=1\linewidth]{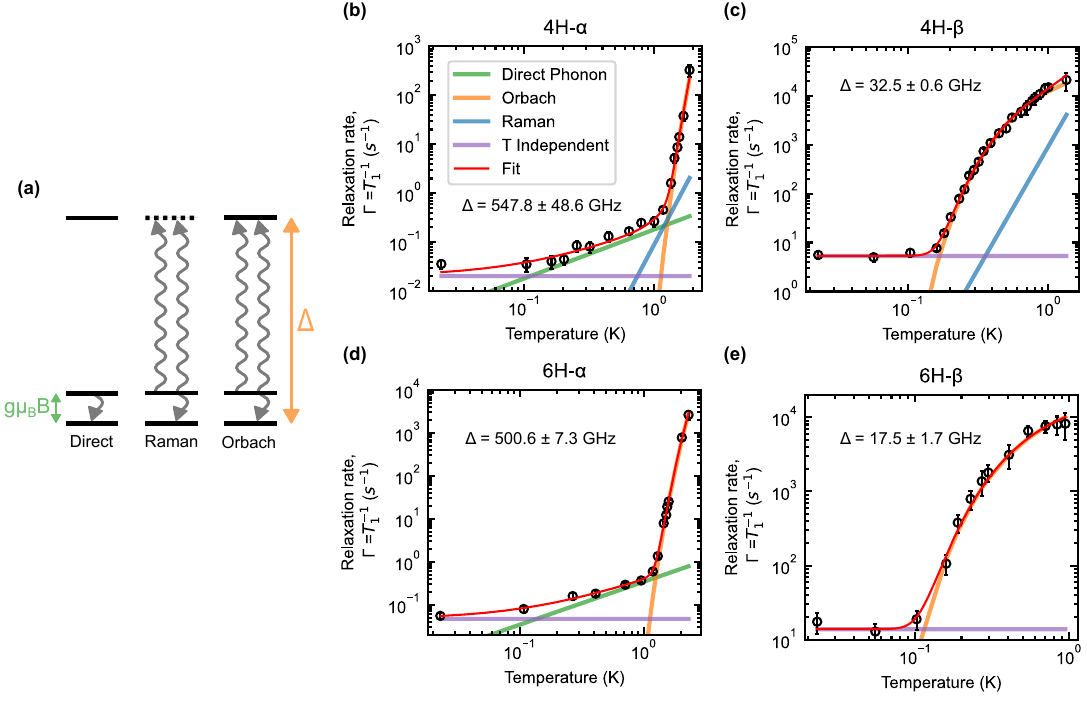}
\caption{\label{fig:Fittings} Spin relaxation processes for $V^{4+}$ in SiC. (a) Diagram of three phonon processes for relaxation mechanisms: Direct phonon process, Raman process, and Orbach process. $g \mu_{B} B$ corresponds to the Zeeman splitting between the ground spin states and $\Delta$ is the energy splitting between the ground and relevant excited state. (b-e) Model fitting of temperature dependent spin relaxation rate for $\alpha$ and $\beta$ sites in 4H- and 6H-SiC. Results show that the Orbach process dominates at higher temperatures and the direct phonon process or temperature independent process dominate at lower temperatures. Inset denotes energy splitting to the relevant excited state in the Orbach process obtained from fitting, matching closely to GS1-GS2 splitting for all measured sites. All reported errors represent 1 SE from the fit, and all error bars represent 1 SD from experimental acquisition.}
\end{figure*}

The strong temperature dependence of the longitudinal spin relaxation times in solids also provides insight into the phonon mediated spin relaxation mechanisms \cite{shrivastava1983theory,abragam1970electron}. These mechanisms include a direct phonon process where one phonon resonant to the two spin sublevels is emitted or absorbed, a Raman process which involves a virtual absorption and emission of two phonons with an energy difference resonant to the two spin sublevels, and the Orbach process where one phonon excites the spin to a higher excited state followed by a phonon emission and relaxation to the opposite spin sublevel \cite{orbach1961spin} (Fig. \ref{fig:Fittings} (a)). The relaxation rates of each process has a temperature dependence where the direct phonon process is proportional to the sample temperature ($1/T_1 \propto T$), while the Raman process can be described by ($1/T_1 \propto T^n$) where n = 5 or 9 for a Kramers doublet \cite{shrivastava1983theory,gilardoni2020spin}, and the Orbach process follows ($1/T_1 \propto e^{-\Delta / k_B T}$) where $\Delta$ is the energy splitting to the phonon-accessible excited state and $k_B$ is the Boltzmann constant. Therefore, the following model can describe the temperature dependent spin relaxation rates:
\begin{equation}\label{E:1}
\begin{aligned}[b]
1/T_1 = A_c + A_D T + A_R T^n + A_Oe^{-\Delta / k_B T}, n = 5, 9
\end{aligned}
\end{equation}
where $A_c$ is the temperature independent constant and $A_D$, $A_R$, and $A_O$ are the coefficients for a direct, Raman, and Orbach processes, respectively.

The temperature dependent relaxation rates ($1/T_1$) from each site are fit to the model above, which illustrates the contribution of each phonon process (Fig. \ref{fig:Fittings} (b)-(e)). At T $<$ 100 mK, a saturation in the measured spin $T_1$ is observed suggesting an effective sample temperature of $\sim$100 mK. While the temperature dependence cannot be probed within this regime, the above-mentioned observation of the prolonged $T_1$ at lower magnetic fields ($A_D \propto B^5$) from the 4H-$\beta$ site implies that the contribution from the direct phonon process is significant at this temperature range \cite{abragam1970electron,dibos2018atomic}. For the $\alpha$ sites, at 100 mK $<$ T $<$ 1K, the relaxation rate is linearly proportional to temperature indicating that the direct phonon process is the dominant spin relaxation mechanism. Finally, at T $>$ 1 K for the $\alpha$ sites and T $>$ 100 mK for the $\beta$ sites, the fittings show that the main phonon process can be attributed to an Orbach process. Critically, the characteristic energy value obtained from the fit, $\Delta$, is consistent with the energy difference to the higher orbital ground state we have measured in previous sections with PLE spectroscopy (Table \ref{tab:Table}). Hence, we conclude that, for all sites measured in our experiments, the Orbach process predominantly involves the upper-branch ground state (GS2) as the excited state of the two-phonon process.

\begin{table}[b]
\caption{\label{tab:Table} Identification of the relevant excited state in the Orbach process. Fitting Eq. (1) to the temperature dependent spin relaxation rates ($1/T_1$), we obtain $\Delta$ from the Orbach term that is consistent with the GS1-GS2 splitting measured from PLE spectroscopy for all measured sites. All reported errors represent 1 SE from the fit.}
\begin{ruledtabular}
\begin{tabular}{ccc}
\textrm{Site}&
\textrm{GS1-GS2 (GHz)}&
\textrm{$\Delta$ (GHz)}\\
\colrule
 4H-$\alpha$ & 530 & 547.8 $\pm$ 48.6 \\  
 4H-$\beta$ & 43 & 32.5 $\pm$ 0.6 \\
 6H-$\alpha$ & 525 & 500.6 $\pm$ 7.3 \\  
 6H-$\beta$ & 25 & 17.5 $\pm$ 1.7 \\
\end{tabular}
\end{ruledtabular}
\end{table}

\begin{figure}[!hbtp]
\centering
\includegraphics[width=\linewidth]{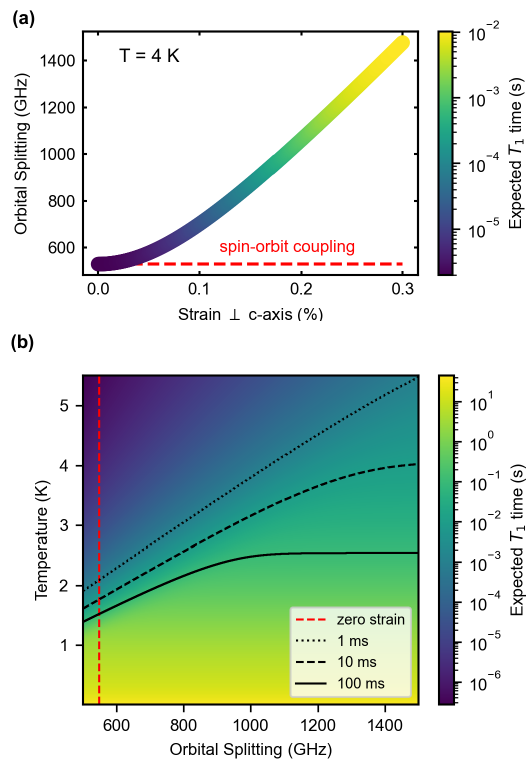}
\caption{\label{fig:Operation Temperature} Strain tuning for increased spin $T_1$ and operation temperature for the 4H-$\alpha$ site. (a) A strain perpendicular to the crystal axis increases the GS1-GS2 orbital splitting, leading to longer spin relaxation times. With a strain of 0.3\%, the GS1-GS2 orbital splitting is increased to approximately 1.5 THz and expected $T_1$ reaches about 10 ms at 4K. (b) Extrapolation of expected spin $T_1$ for 4H-$\alpha$ site at given orbital splitting and temperature. The static strain can enable operation at elevated temperatures that are more accessible with common cryogenic systems.}
\end{figure}

These results are analogous to the spin relaxation mechanisms of group-IV color centers in diamond from which we draw inspiration on methods to elongate the spin $T_1$ of $V^{4+}$ in SiC at elevated temperatures \cite{jahnke2015electron,pingault2017coherent,trusheim2020transform}. Since $V^{4+}$ spins in SiC have demonstrated to be stable near-surface \cite{wolfowicz2020vanadium,cilibrizzi2023ultra}, the use of nanofabricated phononic crystals or using nanoparticles to modify the phononic density of states could be one feasible approach \cite{kuruma2023engineering,klotz2022prolonged}. However, a path that has been more regularly used is the application of static strain to engineer the electron-phonon process and therefore improve both the spin relaxation as well as the spin coherence \cite{meesala2018strain,sohn2018controlling,guo2023microwave,stas2022robust}. Recent theoretical studies on the $V^{4+}$ in SiC has predicted that static strain perpendicular to the c-axis will increase the SO splitting in the ground state, which would lead to an increased $T_1$ at a given temperature \cite{tissot2024strain}. Adopting the model from \cite{tissot2024strain} for the 4H-$\alpha$ site, we calculate the relation between strain and the GS1-GS2 orbital splitting, while extrapolation of our temperature dependent $T_1$ data fitting to Eq.(\ref{E:1}) allows us to estimate the corresponding spin $T_1$ times for a given orbital splitting value. As shown in Fig.\ref{fig:Operation Temperature} (a), a strain of 0.3\% is expected to increase the orbital splitting to $\sim$1.5 THz which would lead to a spin $T_1$ of $\sim$10 ms at 4K. Moreover, with our fittings to Eq.(\ref{E:1}), the expected $T_1$ times can be extrapolated for different temperatures and GS orbital splittings (Fig.\ref{fig:Operation Temperature} (b)), showing that static strain will enable longer spin $T_1$ at higher temperatures. This approach is also applicable to the inequivalent sites of $V^{4+}$ in SiC (see the Supplemental Material \cite{supp}), leading to various options of qubits operating at temperatures that are more accessible with typical cryogenic systems \cite{rosenthal2023microwave,guo2023microwave}.

\section{Discussion and outlook}

In this work, we have demonstrated the potential of $V^{4+}$ in SiC as a telecom solid-state spin qubit system by satisfying the essential requirements of a spin-photon interface, optical spin polarization and readout, as well as showcasing prolonged spin $T_1$ at sub-Kelvin temperatures. Moreover, by extrapolating the different relaxation processes, we propose the utilization of strain tuning to manipulate the orbital levels and mitigate the phonon processes and foresee significant potential in facilitating qubit operations at elevated temperatures. This presents a promising path towards significantly reducing the infrastructure barriers for practical realization of quantum technologies based on $V^{4+}$ in SiC.

Our findings provide compelling motivation for further exploration on the system. For instance, the electron spin coherence of $V^{4+}$ in SiC within this sub-Kelvin temperature regime is an interesting area for additional investigation. Although previous studies at $\sim$2 K have identified the inhomogeneous electric field noise to be the main source limiting $T_2$ when operating near a clock transition, a charge depletion technique could be used to suppress the electric noise and enable studies on other contributions including the central nuclear spin \cite{hendriks2022coherent,onizhuk2021probing}. Additionally, the experimental implementation of nuclear spin polarization will unlock the capability to prepare pure quantum states and to fully utilize the large Hilbert space provided by the electronuclear system \cite{tissot2022nuclear,adambukulam2024hyperfine}. The deterministic access to the intrinsic nuclear spin-7/2 states will not only serve as a long-lived quantum memory but also provide opportunities for generation of multidimensional photonic cluster states for quantum error correction \cite{stas2022robust,parker2023diamond,gross2021designing,michaels2021multidimensional}. Finally, moderate photonic enhancements to aid studies on single defects will provide the platform to test various quantum communication protocols with $V^{4+}$ in SiC.

Going forward, the presence of $V^{4+}$ in the semiconductor host of SiC offers a straightforward path towards device integration. The robust nanofabrication and doping processes available for wafer-scale SiC could enable high-quality, on-chip quantum architectures including nanophotonic, microelectronic, and hybrid spin-mechanical devices \cite{lukin20204h,anderson2019electrical,whiteley2019spin}. Thus, this work propels $V^{4+}$ in SiC to the forefront of telecom wavelength spin qubits for scalable quantum nodes in large-scale quantum communication networks \cite{ecker2024quantum}.

\begin{acknowledgments}
We thank G.Wolfowicz, G. Smith, S. Gupta, N. Delegan, and B. Pingault for helpful discussions and A. Miller (Quantum Opus) for design of experimental mount.

The authors acknowledge primary support from the U.S. Department of Energy, Office of Science; Basic Energy Sciences, Materials Sciences, and Engineering Division (J. A., C.W., N.B., M.T.S., J.Z., F. J.H., and D.D.A.) for project conceptualization and experimental lead.  Additional support for experimental setup development was provided by the Q-NEXT Quantum Center, a U.S. Department of Energy, Office of Science, National Quantum Information Science Research Center (A.M.D.) under Award Number DE-FOA-0002253. B.T. and G.B. acknowledge funding from the German Federal Ministry of Education and Research (BMBF) under the Grant Agreement No. 13N16212 (SPINNING).

\end{acknowledgments}

\appendix

\section{Samples}

Both 4H-SiC and 6H-SiC samples are 500 $\mu$m thick vanadium compensated semi-insulating SiC wafers commercially purchased from II-VI materials. Secondary ion mass spectroscopy measurements (Eurofins EAG Laboratories) are carried out on sample pieces diced from the same wafer and reports the following concentrations of dopants in 4H-SiC (V: 1.3e17$/cm^3$, N: 7e16$/cm^3$, B: 3.1e15$/cm^3$, Al: 2.2e14$/cm^3$) and 6H-SiC (V: 1.4e17$/cm^3$, N: 3.7e16$/cm^3$, B: 6.2e15$/cm^3$, Al: 1.6e14$/cm^3$). No annealing step is performed for the samples.

\section{Optical spectroscopy}

All optical spectroscopy measurements are performed in a dilution refrigerator setup (see the Supplemental Material \cite{supp}) with the 4H- and 6H-SiC samples mounted on the mixing plate with temperatures between 22 mK and 4 K. The samples are excited with a narrow linewidth, tunable O-band laser fiber coupled into the dilution refrigerator where the fiber ferrule is aligned parallel to the crystal axis (c-axis) of the sample. As a result, the resonant laser used for the excitation of the optical transitions is mostly polarized perpendicular to the c-axis. The laser output is pulsed with an acousto-optic modulator and the resonant excitation is about $\sim$75 nW at sample, unless stated otherwise. For the 4H-SiC sample, a 405 nm laser is used to compensate for resonant laser-induced ionization and laser power for most experiments is about $\sim$400 nW at sample. PL emission is only collected from the phonon sideband by using a set of optical filters to spectrally filter out the laser reflection. Finally, the PL is detected with a superconducting nanowire single-photon detector. More details on the experimental setup is provided in the Supplemental Material \cite{supp}.

\section{Optical spin polarization and hole burning recovery measurements}

For optical spin polarization and hole burning recovery measurements, an external magnetic field of 250 mT parallel to the SiC c-axis is applied using a superconducting vector magnet. For the 4H-SiC, a 405 nm laser pulse with $\sim$400 nW is applied for $>$100 s in between measurements to assure the vanadium ensemble is in the desired $V^{4+}$ charge state while avoiding raising the sample temperature. This charge repump is not necessary and omitted for 6H-SiC. The duration of the resonant laser pulses are set to 2 ms which are many orders of magnitude shorter than the photoionization timescales while sufficiently long compared to the polarization timescales. Therefore, the charge effects are negligible while population are successfully accumulated to the other spin sublevel.

\nocite{*}

\bibliographystyle{apsrev4-2}
\bibliography{main}

\begin{thebibliography}{59}%
\makeatletter
\providecommand \@ifxundefined [1]{%
 \@ifx{#1\undefined}
}%
\providecommand \@ifnum [1]{%
 \ifnum #1\expandafter \@firstoftwo
 \else \expandafter \@secondoftwo
 \fi
}%
\providecommand \@ifx [1]{%
 \ifx #1\expandafter \@firstoftwo
 \else \expandafter \@secondoftwo
 \fi
}%
\providecommand \natexlab [1]{#1}%
\providecommand \enquote  [1]{``#1''}%
\providecommand \bibnamefont  [1]{#1}%
\providecommand \bibfnamefont [1]{#1}%
\providecommand \citenamefont [1]{#1}%
\providecommand \href@noop [0]{\@secondoftwo}%
\providecommand \href [0]{\begingroup \@sanitize@url \@href}%
\providecommand \@href[1]{\@@startlink{#1}\@@href}%
\providecommand \@@href[1]{\endgroup#1\@@endlink}%
\providecommand \@sanitize@url [0]{\catcode `\\12\catcode `\$12\catcode `\&12\catcode `\#12\catcode `\^12\catcode `\_12\catcode `\%12\relax}%
\providecommand \@@startlink[1]{}%
\providecommand \@@endlink[0]{}%
\providecommand \url  [0]{\begingroup\@sanitize@url \@url }%
\providecommand \@url [1]{\endgroup\@href {#1}{\urlprefix }}%
\providecommand \urlprefix  [0]{URL }%
\providecommand \Eprint [0]{\href }%
\providecommand \doibase [0]{https://doi.org/}%
\providecommand \selectlanguage [0]{\@gobble}%
\providecommand \bibinfo  [0]{\@secondoftwo}%
\providecommand \bibfield  [0]{\@secondoftwo}%
\providecommand \translation [1]{[#1]}%
\providecommand \BibitemOpen [0]{}%
\providecommand \bibitemStop [0]{}%
\providecommand \bibitemNoStop [0]{.\EOS\space}%
\providecommand \EOS [0]{\spacefactor3000\relax}%
\providecommand \BibitemShut  [1]{\csname bibitem#1\endcsname}%
\let\auto@bib@innerbib\@empty
\bibitem [{\citenamefont {Wolfowicz}\ \emph {et~al.}(2021)\citenamefont {Wolfowicz}, \citenamefont {Heremans}, \citenamefont {Anderson}, \citenamefont {Kanai}, \citenamefont {Seo}, \citenamefont {Gali}, \citenamefont {Galli},\ and\ \citenamefont {Awschalom}}]{wolfowicz2021quantum}%
  \BibitemOpen
  \bibfield  {author} {\bibinfo {author} {\bibfnamefont {G.}~\bibnamefont {Wolfowicz}}, \bibinfo {author} {\bibfnamefont {F.~J.}\ \bibnamefont {Heremans}}, \bibinfo {author} {\bibfnamefont {C.~P.}\ \bibnamefont {Anderson}}, \bibinfo {author} {\bibfnamefont {S.}~\bibnamefont {Kanai}}, \bibinfo {author} {\bibfnamefont {H.}~\bibnamefont {Seo}}, \bibinfo {author} {\bibfnamefont {A.}~\bibnamefont {Gali}}, \bibinfo {author} {\bibfnamefont {G.}~\bibnamefont {Galli}},\ and\ \bibinfo {author} {\bibfnamefont {D.~D.}\ \bibnamefont {Awschalom}},\ }\href@noop {} {\bibfield  {journal} {\bibinfo  {journal} {Nature Reviews Materials}\ }\textbf {\bibinfo {volume} {6}},\ \bibinfo {pages} {906} (\bibinfo {year} {2021})}\BibitemShut {NoStop}%
\bibitem [{\citenamefont {Hermans}\ \emph {et~al.}(2022)\citenamefont {Hermans}, \citenamefont {Pompili}, \citenamefont {Beukers}, \citenamefont {Baier}, \citenamefont {Borregaard},\ and\ \citenamefont {Hanson}}]{hermans2022qubit}%
  \BibitemOpen
  \bibfield  {author} {\bibinfo {author} {\bibfnamefont {S.}~\bibnamefont {Hermans}}, \bibinfo {author} {\bibfnamefont {M.}~\bibnamefont {Pompili}}, \bibinfo {author} {\bibfnamefont {H.}~\bibnamefont {Beukers}}, \bibinfo {author} {\bibfnamefont {S.}~\bibnamefont {Baier}}, \bibinfo {author} {\bibfnamefont {J.}~\bibnamefont {Borregaard}},\ and\ \bibinfo {author} {\bibfnamefont {R.}~\bibnamefont {Hanson}},\ }\href@noop {} {\bibfield  {journal} {\bibinfo  {journal} {Nature}\ }\textbf {\bibinfo {volume} {605}},\ \bibinfo {pages} {663} (\bibinfo {year} {2022})}\BibitemShut {NoStop}%
\bibitem [{\citenamefont {Knaut}\ \emph {et~al.}(2023)\citenamefont {Knaut}, \citenamefont {Suleymanzade}, \citenamefont {Wei}, \citenamefont {Assumpcao}, \citenamefont {Stas}, \citenamefont {Huan}, \citenamefont {Machielse}, \citenamefont {Knall}, \citenamefont {Sutula}, \citenamefont {Baranes} \emph {et~al.}}]{knaut2023entanglement}%
  \BibitemOpen
  \bibfield  {author} {\bibinfo {author} {\bibfnamefont {C.~M.}\ \bibnamefont {Knaut}}, \bibinfo {author} {\bibfnamefont {A.}~\bibnamefont {Suleymanzade}}, \bibinfo {author} {\bibfnamefont {Y.-C.}\ \bibnamefont {Wei}}, \bibinfo {author} {\bibfnamefont {D.~R.}\ \bibnamefont {Assumpcao}}, \bibinfo {author} {\bibfnamefont {P.-J.}\ \bibnamefont {Stas}}, \bibinfo {author} {\bibfnamefont {Y.~Q.}\ \bibnamefont {Huan}}, \bibinfo {author} {\bibfnamefont {B.}~\bibnamefont {Machielse}}, \bibinfo {author} {\bibfnamefont {E.~N.}\ \bibnamefont {Knall}}, \bibinfo {author} {\bibfnamefont {M.}~\bibnamefont {Sutula}}, \bibinfo {author} {\bibfnamefont {G.}~\bibnamefont {Baranes}}, \emph {et~al.},\ }\href@noop {} {\bibfield  {journal} {\bibinfo  {journal} {arXiv preprint arXiv:2310.01316}\ } (\bibinfo {year} {2023})}\BibitemShut {NoStop}%
\bibitem [{\citenamefont {Dibos}\ \emph {et~al.}(2018)\citenamefont {Dibos}, \citenamefont {Raha}, \citenamefont {Phenicie},\ and\ \citenamefont {Thompson}}]{dibos2018atomic}%
  \BibitemOpen
  \bibfield  {author} {\bibinfo {author} {\bibfnamefont {A.}~\bibnamefont {Dibos}}, \bibinfo {author} {\bibfnamefont {M.}~\bibnamefont {Raha}}, \bibinfo {author} {\bibfnamefont {C.}~\bibnamefont {Phenicie}},\ and\ \bibinfo {author} {\bibfnamefont {J.~D.}\ \bibnamefont {Thompson}},\ }\href@noop {} {\bibfield  {journal} {\bibinfo  {journal} {Physical Review Letters}\ }\textbf {\bibinfo {volume} {120}},\ \bibinfo {pages} {243601} (\bibinfo {year} {2018})}\BibitemShut {NoStop}%
\bibitem [{\citenamefont {Ourari}\ \emph {et~al.}(2023)\citenamefont {Ourari}, \citenamefont {Dusanowski}, \citenamefont {Horvath}, \citenamefont {Uysal}, \citenamefont {Phenicie}, \citenamefont {Stevenson}, \citenamefont {Raha}, \citenamefont {Chen}, \citenamefont {Cava}, \citenamefont {de~Leon} \emph {et~al.}}]{ourari2023indistinguishable}%
  \BibitemOpen
  \bibfield  {author} {\bibinfo {author} {\bibfnamefont {S.}~\bibnamefont {Ourari}}, \bibinfo {author} {\bibfnamefont {{\L}.}~\bibnamefont {Dusanowski}}, \bibinfo {author} {\bibfnamefont {S.~P.}\ \bibnamefont {Horvath}}, \bibinfo {author} {\bibfnamefont {M.~T.}\ \bibnamefont {Uysal}}, \bibinfo {author} {\bibfnamefont {C.~M.}\ \bibnamefont {Phenicie}}, \bibinfo {author} {\bibfnamefont {P.}~\bibnamefont {Stevenson}}, \bibinfo {author} {\bibfnamefont {M.}~\bibnamefont {Raha}}, \bibinfo {author} {\bibfnamefont {S.}~\bibnamefont {Chen}}, \bibinfo {author} {\bibfnamefont {R.~J.}\ \bibnamefont {Cava}}, \bibinfo {author} {\bibfnamefont {N.~P.}\ \bibnamefont {de~Leon}}, \emph {et~al.},\ }\href@noop {} {\bibfield  {journal} {\bibinfo  {journal} {Nature}\ }\textbf {\bibinfo {volume} {620}},\ \bibinfo {pages} {977} (\bibinfo {year} {2023})}\BibitemShut {NoStop}%
\bibitem [{\citenamefont {Kurkjian}\ \emph {et~al.}(2021)\citenamefont {Kurkjian}, \citenamefont {Higginbottom}, \citenamefont {Chartrand}, \citenamefont {MacQuarrie}, \citenamefont {Klein}, \citenamefont {Lee-Hone}, \citenamefont {Stacho}, \citenamefont {Bowness}, \citenamefont {Bergeron}, \citenamefont {DeAbreu} \emph {et~al.}}]{kurkjian2021optical}%
  \BibitemOpen
  \bibfield  {author} {\bibinfo {author} {\bibfnamefont {A.}~\bibnamefont {Kurkjian}}, \bibinfo {author} {\bibfnamefont {D.}~\bibnamefont {Higginbottom}}, \bibinfo {author} {\bibfnamefont {C.}~\bibnamefont {Chartrand}}, \bibinfo {author} {\bibfnamefont {E.}~\bibnamefont {MacQuarrie}}, \bibinfo {author} {\bibfnamefont {J.}~\bibnamefont {Klein}}, \bibinfo {author} {\bibfnamefont {N.}~\bibnamefont {Lee-Hone}}, \bibinfo {author} {\bibfnamefont {J.}~\bibnamefont {Stacho}}, \bibinfo {author} {\bibfnamefont {C.}~\bibnamefont {Bowness}}, \bibinfo {author} {\bibfnamefont {L.}~\bibnamefont {Bergeron}}, \bibinfo {author} {\bibfnamefont {A.}~\bibnamefont {DeAbreu}}, \emph {et~al.},\ }\href@noop {} {\bibfield  {journal} {\bibinfo  {journal} {arXiv preprint arXiv:2103.07580}\ } (\bibinfo {year} {2021})}\BibitemShut {NoStop}%
\bibitem [{\citenamefont {Dr{\'e}au}\ \emph {et~al.}(2018)\citenamefont {Dr{\'e}au}, \citenamefont {Tchebotareva}, \citenamefont {El~Mahdaoui}, \citenamefont {Bonato},\ and\ \citenamefont {Hanson}}]{dreau2018quantum}%
  \BibitemOpen
  \bibfield  {author} {\bibinfo {author} {\bibfnamefont {A.}~\bibnamefont {Dr{\'e}au}}, \bibinfo {author} {\bibfnamefont {A.}~\bibnamefont {Tchebotareva}}, \bibinfo {author} {\bibfnamefont {A.}~\bibnamefont {El~Mahdaoui}}, \bibinfo {author} {\bibfnamefont {C.}~\bibnamefont {Bonato}},\ and\ \bibinfo {author} {\bibfnamefont {R.}~\bibnamefont {Hanson}},\ }\href@noop {} {\bibfield  {journal} {\bibinfo  {journal} {Physical Review Applied}\ }\textbf {\bibinfo {volume} {9}},\ \bibinfo {pages} {064031} (\bibinfo {year} {2018})}\BibitemShut {NoStop}%
\bibitem [{\citenamefont {Bersin}\ \emph {et~al.}(2024)\citenamefont {Bersin}, \citenamefont {Sutula}, \citenamefont {Huan}, \citenamefont {Suleymanzade}, \citenamefont {Assumpcao}, \citenamefont {Wei}, \citenamefont {Stas}, \citenamefont {Knaut}, \citenamefont {Knall}, \citenamefont {Langrock} \emph {et~al.}}]{bersin2024telecom}%
  \BibitemOpen
  \bibfield  {author} {\bibinfo {author} {\bibfnamefont {E.}~\bibnamefont {Bersin}}, \bibinfo {author} {\bibfnamefont {M.}~\bibnamefont {Sutula}}, \bibinfo {author} {\bibfnamefont {Y.~Q.}\ \bibnamefont {Huan}}, \bibinfo {author} {\bibfnamefont {A.}~\bibnamefont {Suleymanzade}}, \bibinfo {author} {\bibfnamefont {D.~R.}\ \bibnamefont {Assumpcao}}, \bibinfo {author} {\bibfnamefont {Y.-C.}\ \bibnamefont {Wei}}, \bibinfo {author} {\bibfnamefont {P.-J.}\ \bibnamefont {Stas}}, \bibinfo {author} {\bibfnamefont {C.~M.}\ \bibnamefont {Knaut}}, \bibinfo {author} {\bibfnamefont {E.~N.}\ \bibnamefont {Knall}}, \bibinfo {author} {\bibfnamefont {C.}~\bibnamefont {Langrock}}, \emph {et~al.},\ }\href@noop {} {\bibfield  {journal} {\bibinfo  {journal} {PRX Quantum}\ }\textbf {\bibinfo {volume} {5}},\ \bibinfo {pages} {010303} (\bibinfo {year} {2024})}\BibitemShut {NoStop}%
\bibitem [{\citenamefont {Spindlberger}\ \emph {et~al.}(2019)\citenamefont {Spindlberger}, \citenamefont {Cs{\'o}r{\'e}}, \citenamefont {Thiering}, \citenamefont {Putz}, \citenamefont {Karhu}, \citenamefont {Hassan}, \citenamefont {Son}, \citenamefont {Fromherz}, \citenamefont {Gali},\ and\ \citenamefont {Trupke}}]{spindlberger2019optical}%
  \BibitemOpen
  \bibfield  {author} {\bibinfo {author} {\bibfnamefont {L.}~\bibnamefont {Spindlberger}}, \bibinfo {author} {\bibfnamefont {A.}~\bibnamefont {Cs{\'o}r{\'e}}}, \bibinfo {author} {\bibfnamefont {G.}~\bibnamefont {Thiering}}, \bibinfo {author} {\bibfnamefont {S.}~\bibnamefont {Putz}}, \bibinfo {author} {\bibfnamefont {R.}~\bibnamefont {Karhu}}, \bibinfo {author} {\bibfnamefont {J.~U.}\ \bibnamefont {Hassan}}, \bibinfo {author} {\bibfnamefont {N.}~\bibnamefont {Son}}, \bibinfo {author} {\bibfnamefont {T.}~\bibnamefont {Fromherz}}, \bibinfo {author} {\bibfnamefont {A.}~\bibnamefont {Gali}},\ and\ \bibinfo {author} {\bibfnamefont {M.}~\bibnamefont {Trupke}},\ }\href@noop {} {\bibfield  {journal} {\bibinfo  {journal} {Physical Review Applied}\ }\textbf {\bibinfo {volume} {12}},\ \bibinfo {pages} {014015} (\bibinfo {year} {2019})}\BibitemShut {NoStop}%
\bibitem [{\citenamefont {Wolfowicz}\ \emph {et~al.}(2020)\citenamefont {Wolfowicz}, \citenamefont {Anderson}, \citenamefont {Diler}, \citenamefont {Poluektov}, \citenamefont {Heremans},\ and\ \citenamefont {Awschalom}}]{wolfowicz2020vanadium}%
  \BibitemOpen
  \bibfield  {author} {\bibinfo {author} {\bibfnamefont {G.}~\bibnamefont {Wolfowicz}}, \bibinfo {author} {\bibfnamefont {C.~P.}\ \bibnamefont {Anderson}}, \bibinfo {author} {\bibfnamefont {B.}~\bibnamefont {Diler}}, \bibinfo {author} {\bibfnamefont {O.~G.}\ \bibnamefont {Poluektov}}, \bibinfo {author} {\bibfnamefont {F.~J.}\ \bibnamefont {Heremans}},\ and\ \bibinfo {author} {\bibfnamefont {D.~D.}\ \bibnamefont {Awschalom}},\ }\href@noop {} {\bibfield  {journal} {\bibinfo  {journal} {Science Advances}\ }\textbf {\bibinfo {volume} {6}},\ \bibinfo {pages} {eaaz1192} (\bibinfo {year} {2020})}\BibitemShut {NoStop}%
\bibitem [{\citenamefont {Cilibrizzi}\ \emph {et~al.}(2023)\citenamefont {Cilibrizzi}, \citenamefont {Arshad}, \citenamefont {Tissot}, \citenamefont {Son}, \citenamefont {Ivanov}, \citenamefont {Astner}, \citenamefont {Koller}, \citenamefont {Ghezellou}, \citenamefont {Ul-Hassan}, \citenamefont {White} \emph {et~al.}}]{cilibrizzi2023ultra}%
  \BibitemOpen
  \bibfield  {author} {\bibinfo {author} {\bibfnamefont {P.}~\bibnamefont {Cilibrizzi}}, \bibinfo {author} {\bibfnamefont {M.~J.}\ \bibnamefont {Arshad}}, \bibinfo {author} {\bibfnamefont {B.}~\bibnamefont {Tissot}}, \bibinfo {author} {\bibfnamefont {N.~T.}\ \bibnamefont {Son}}, \bibinfo {author} {\bibfnamefont {I.~G.}\ \bibnamefont {Ivanov}}, \bibinfo {author} {\bibfnamefont {T.}~\bibnamefont {Astner}}, \bibinfo {author} {\bibfnamefont {P.}~\bibnamefont {Koller}}, \bibinfo {author} {\bibfnamefont {M.}~\bibnamefont {Ghezellou}}, \bibinfo {author} {\bibfnamefont {J.}~\bibnamefont {Ul-Hassan}}, \bibinfo {author} {\bibfnamefont {D.}~\bibnamefont {White}}, \emph {et~al.},\ }\href@noop {} {\bibfield  {journal} {\bibinfo  {journal} {Nature Communications}\ }\textbf {\bibinfo {volume} {14}},\ \bibinfo {pages} {8448} (\bibinfo {year} {2023})}\BibitemShut {NoStop}%
\bibitem [{\citenamefont {Astner}\ \emph {et~al.}(2022)\citenamefont {Astner}, \citenamefont {Koller}, \citenamefont {Gilardoni}, \citenamefont {Hendriks}, \citenamefont {Son}, \citenamefont {Ivanov}, \citenamefont {Hassan}, \citenamefont {van~der Wal},\ and\ \citenamefont {Trupke}}]{astner2022vanadium}%
  \BibitemOpen
  \bibfield  {author} {\bibinfo {author} {\bibfnamefont {T.}~\bibnamefont {Astner}}, \bibinfo {author} {\bibfnamefont {P.}~\bibnamefont {Koller}}, \bibinfo {author} {\bibfnamefont {C.~M.}\ \bibnamefont {Gilardoni}}, \bibinfo {author} {\bibfnamefont {J.}~\bibnamefont {Hendriks}}, \bibinfo {author} {\bibfnamefont {N.}~\bibnamefont {Son}}, \bibinfo {author} {\bibfnamefont {I.}~\bibnamefont {Ivanov}}, \bibinfo {author} {\bibfnamefont {J.}~\bibnamefont {Hassan}}, \bibinfo {author} {\bibfnamefont {C.}~\bibnamefont {van~der Wal}},\ and\ \bibinfo {author} {\bibfnamefont {M.}~\bibnamefont {Trupke}},\ }\href@noop {} {\bibfield  {journal} {\bibinfo  {journal} {arXiv preprint arXiv:2206.06240}\ } (\bibinfo {year} {2022})}\BibitemShut {NoStop}%
\bibitem [{\citenamefont {Hendriks}\ \emph {et~al.}(2022)\citenamefont {Hendriks}, \citenamefont {Gilardoni}, \citenamefont {Adambukulam}, \citenamefont {Laucht},\ and\ \citenamefont {van~der Wal}}]{hendriks2022coherent}%
  \BibitemOpen
  \bibfield  {author} {\bibinfo {author} {\bibfnamefont {J.}~\bibnamefont {Hendriks}}, \bibinfo {author} {\bibfnamefont {C.~M.}\ \bibnamefont {Gilardoni}}, \bibinfo {author} {\bibfnamefont {C.}~\bibnamefont {Adambukulam}}, \bibinfo {author} {\bibfnamefont {A.}~\bibnamefont {Laucht}},\ and\ \bibinfo {author} {\bibfnamefont {C.~H.}\ \bibnamefont {van~der Wal}},\ }\href@noop {} {\bibfield  {journal} {\bibinfo  {journal} {arXiv preprint arXiv:2210.09942}\ } (\bibinfo {year} {2022})}\BibitemShut {NoStop}%
\bibitem [{\citenamefont {Lukin}\ \emph {et~al.}(2020)\citenamefont {Lukin}, \citenamefont {Dory}, \citenamefont {Guidry}, \citenamefont {Yang}, \citenamefont {Mishra}, \citenamefont {Trivedi}, \citenamefont {Radulaski}, \citenamefont {Sun}, \citenamefont {Vercruysse}, \citenamefont {Ahn} \emph {et~al.}}]{lukin20204h}%
  \BibitemOpen
  \bibfield  {author} {\bibinfo {author} {\bibfnamefont {D.~M.}\ \bibnamefont {Lukin}}, \bibinfo {author} {\bibfnamefont {C.}~\bibnamefont {Dory}}, \bibinfo {author} {\bibfnamefont {M.~A.}\ \bibnamefont {Guidry}}, \bibinfo {author} {\bibfnamefont {K.~Y.}\ \bibnamefont {Yang}}, \bibinfo {author} {\bibfnamefont {S.~D.}\ \bibnamefont {Mishra}}, \bibinfo {author} {\bibfnamefont {R.}~\bibnamefont {Trivedi}}, \bibinfo {author} {\bibfnamefont {M.}~\bibnamefont {Radulaski}}, \bibinfo {author} {\bibfnamefont {S.}~\bibnamefont {Sun}}, \bibinfo {author} {\bibfnamefont {D.}~\bibnamefont {Vercruysse}}, \bibinfo {author} {\bibfnamefont {G.~H.}\ \bibnamefont {Ahn}}, \emph {et~al.},\ }\href@noop {} {\bibfield  {journal} {\bibinfo  {journal} {Nature Photonics}\ }\textbf {\bibinfo {volume} {14}},\ \bibinfo {pages} {330} (\bibinfo {year} {2020})}\BibitemShut {NoStop}%
\bibitem [{\citenamefont {Anderson}\ \emph {et~al.}(2019)\citenamefont {Anderson}, \citenamefont {Bourassa}, \citenamefont {Miao}, \citenamefont {Wolfowicz}, \citenamefont {Mintun}, \citenamefont {Crook}, \citenamefont {Abe}, \citenamefont {Ul~Hassan}, \citenamefont {Son}, \citenamefont {Ohshima} \emph {et~al.}}]{anderson2019electrical}%
  \BibitemOpen
  \bibfield  {author} {\bibinfo {author} {\bibfnamefont {C.~P.}\ \bibnamefont {Anderson}}, \bibinfo {author} {\bibfnamefont {A.}~\bibnamefont {Bourassa}}, \bibinfo {author} {\bibfnamefont {K.~C.}\ \bibnamefont {Miao}}, \bibinfo {author} {\bibfnamefont {G.}~\bibnamefont {Wolfowicz}}, \bibinfo {author} {\bibfnamefont {P.~J.}\ \bibnamefont {Mintun}}, \bibinfo {author} {\bibfnamefont {A.~L.}\ \bibnamefont {Crook}}, \bibinfo {author} {\bibfnamefont {H.}~\bibnamefont {Abe}}, \bibinfo {author} {\bibfnamefont {J.}~\bibnamefont {Ul~Hassan}}, \bibinfo {author} {\bibfnamefont {N.~T.}\ \bibnamefont {Son}}, \bibinfo {author} {\bibfnamefont {T.}~\bibnamefont {Ohshima}}, \emph {et~al.},\ }\href@noop {} {\bibfield  {journal} {\bibinfo  {journal} {Science}\ }\textbf {\bibinfo {volume} {366}},\ \bibinfo {pages} {1225} (\bibinfo {year} {2019})}\BibitemShut {NoStop}%
\bibitem [{\citenamefont {Whiteley}\ \emph {et~al.}(2019)\citenamefont {Whiteley}, \citenamefont {Wolfowicz}, \citenamefont {Anderson}, \citenamefont {Bourassa}, \citenamefont {Ma}, \citenamefont {Ye}, \citenamefont {Koolstra}, \citenamefont {Satzinger}, \citenamefont {Holt}, \citenamefont {Heremans} \emph {et~al.}}]{whiteley2019spin}%
  \BibitemOpen
  \bibfield  {author} {\bibinfo {author} {\bibfnamefont {S.~J.}\ \bibnamefont {Whiteley}}, \bibinfo {author} {\bibfnamefont {G.}~\bibnamefont {Wolfowicz}}, \bibinfo {author} {\bibfnamefont {C.~P.}\ \bibnamefont {Anderson}}, \bibinfo {author} {\bibfnamefont {A.}~\bibnamefont {Bourassa}}, \bibinfo {author} {\bibfnamefont {H.}~\bibnamefont {Ma}}, \bibinfo {author} {\bibfnamefont {M.}~\bibnamefont {Ye}}, \bibinfo {author} {\bibfnamefont {G.}~\bibnamefont {Koolstra}}, \bibinfo {author} {\bibfnamefont {K.~J.}\ \bibnamefont {Satzinger}}, \bibinfo {author} {\bibfnamefont {M.~V.}\ \bibnamefont {Holt}}, \bibinfo {author} {\bibfnamefont {F.~J.}\ \bibnamefont {Heremans}}, \emph {et~al.},\ }\href@noop {} {\bibfield  {journal} {\bibinfo  {journal} {Nature Physics}\ }\textbf {\bibinfo {volume} {15}},\ \bibinfo {pages} {490} (\bibinfo {year} {2019})}\BibitemShut {NoStop}%
\bibitem [{\citenamefont {Kunzer}\ \emph {et~al.}(1993)\citenamefont {Kunzer}, \citenamefont {M{\"u}ller},\ and\ \citenamefont {Kaufmann}}]{kunzer1993magnetic}%
  \BibitemOpen
  \bibfield  {author} {\bibinfo {author} {\bibfnamefont {M.}~\bibnamefont {Kunzer}}, \bibinfo {author} {\bibfnamefont {H.}~\bibnamefont {M{\"u}ller}},\ and\ \bibinfo {author} {\bibfnamefont {U.}~\bibnamefont {Kaufmann}},\ }\href@noop {} {\bibfield  {journal} {\bibinfo  {journal} {Physical Review B}\ }\textbf {\bibinfo {volume} {48}},\ \bibinfo {pages} {10846} (\bibinfo {year} {1993})}\BibitemShut {NoStop}%
\bibitem [{\citenamefont {Kaufmann}\ \emph {et~al.}(1997)\citenamefont {Kaufmann}, \citenamefont {D{\"o}rnen},\ and\ \citenamefont {Ham}}]{kaufmann1997crystal}%
  \BibitemOpen
  \bibfield  {author} {\bibinfo {author} {\bibfnamefont {B.}~\bibnamefont {Kaufmann}}, \bibinfo {author} {\bibfnamefont {A.}~\bibnamefont {D{\"o}rnen}},\ and\ \bibinfo {author} {\bibfnamefont {F.}~\bibnamefont {Ham}},\ }\href@noop {} {\bibfield  {journal} {\bibinfo  {journal} {Physical Review B}\ }\textbf {\bibinfo {volume} {55}},\ \bibinfo {pages} {13009} (\bibinfo {year} {1997})}\BibitemShut {NoStop}%
\bibitem [{\citenamefont {Baur}\ \emph {et~al.}(1997)\citenamefont {Baur}, \citenamefont {Kunzer},\ and\ \citenamefont {Schneider}}]{baur1997transition}%
  \BibitemOpen
  \bibfield  {author} {\bibinfo {author} {\bibfnamefont {J.}~\bibnamefont {Baur}}, \bibinfo {author} {\bibfnamefont {M.}~\bibnamefont {Kunzer}},\ and\ \bibinfo {author} {\bibfnamefont {J.}~\bibnamefont {Schneider}},\ }\href@noop {} {\bibfield  {journal} {\bibinfo  {journal} {physica status solidi (a)}\ }\textbf {\bibinfo {volume} {162}},\ \bibinfo {pages} {153} (\bibinfo {year} {1997})}\BibitemShut {NoStop}%
\bibitem [{\citenamefont {Tissot}\ and\ \citenamefont {Burkard}(2021{\natexlab{a}})}]{tissot2021spin}%
  \BibitemOpen
  \bibfield  {author} {\bibinfo {author} {\bibfnamefont {B.}~\bibnamefont {Tissot}}\ and\ \bibinfo {author} {\bibfnamefont {G.}~\bibnamefont {Burkard}},\ }\href@noop {} {\bibfield  {journal} {\bibinfo  {journal} {Physical Review B}\ }\textbf {\bibinfo {volume} {103}},\ \bibinfo {pages} {064106} (\bibinfo {year} {2021}{\natexlab{a}})}\BibitemShut {NoStop}%
\bibitem [{\citenamefont {Tissot}\ \emph {et~al.}(2022)\citenamefont {Tissot}, \citenamefont {Trupke}, \citenamefont {Koller}, \citenamefont {Astner},\ and\ \citenamefont {Burkard}}]{tissot2022nuclear}%
  \BibitemOpen
  \bibfield  {author} {\bibinfo {author} {\bibfnamefont {B.}~\bibnamefont {Tissot}}, \bibinfo {author} {\bibfnamefont {M.}~\bibnamefont {Trupke}}, \bibinfo {author} {\bibfnamefont {P.}~\bibnamefont {Koller}}, \bibinfo {author} {\bibfnamefont {T.}~\bibnamefont {Astner}},\ and\ \bibinfo {author} {\bibfnamefont {G.}~\bibnamefont {Burkard}},\ }\href@noop {} {\bibfield  {journal} {\bibinfo  {journal} {Physical Review Research}\ }\textbf {\bibinfo {volume} {4}},\ \bibinfo {pages} {033107} (\bibinfo {year} {2022})}\BibitemShut {NoStop}%
\bibitem [{\citenamefont {Csore}\ and\ \citenamefont {Gali}(2020)}]{csore2020ab}%
  \BibitemOpen
  \bibfield  {author} {\bibinfo {author} {\bibfnamefont {A.}~\bibnamefont {Csore}}\ and\ \bibinfo {author} {\bibfnamefont {A.}~\bibnamefont {Gali}},\ }\href@noop {} {\bibfield  {journal} {\bibinfo  {journal} {Physical Review B}\ }\textbf {\bibinfo {volume} {102}},\ \bibinfo {pages} {241201} (\bibinfo {year} {2020})}\BibitemShut {NoStop}%
\bibitem [{sup()}]{supp}%
  \BibitemOpen
  \href@noop {} {}\bibinfo {note} {See Supplemental Material at [URL will be inserted by publisher] for providing further details on experimental setup, assignment of sites and irreducible representations, origin of spin non-conserving orbital transitions, as well as experimental data regarding the temperature dependent PLE measurements, charge measurements, magnetic field dependent spin relaxation rates, and extrapolation of expected spin T1 for additional sites.}\BibitemShut {Stop}%
\bibitem [{\citenamefont {Green}\ \emph {et~al.}(2017)\citenamefont {Green}, \citenamefont {Mottishaw}, \citenamefont {Breeze}, \citenamefont {Edmonds}, \citenamefont {D’Haenens-Johansson}, \citenamefont {Doherty}, \citenamefont {Williams}, \citenamefont {Twitchen},\ and\ \citenamefont {Newton}}]{green2017neutral}%
  \BibitemOpen
  \bibfield  {author} {\bibinfo {author} {\bibfnamefont {B.}~\bibnamefont {Green}}, \bibinfo {author} {\bibfnamefont {S.}~\bibnamefont {Mottishaw}}, \bibinfo {author} {\bibfnamefont {B.}~\bibnamefont {Breeze}}, \bibinfo {author} {\bibfnamefont {A.}~\bibnamefont {Edmonds}}, \bibinfo {author} {\bibfnamefont {U.}~\bibnamefont {D’Haenens-Johansson}}, \bibinfo {author} {\bibfnamefont {M.}~\bibnamefont {Doherty}}, \bibinfo {author} {\bibfnamefont {S.}~\bibnamefont {Williams}}, \bibinfo {author} {\bibfnamefont {D.}~\bibnamefont {Twitchen}},\ and\ \bibinfo {author} {\bibfnamefont {M.}~\bibnamefont {Newton}},\ }\href@noop {} {\bibfield  {journal} {\bibinfo  {journal} {Physical Review Letters}\ }\textbf {\bibinfo {volume} {119}},\ \bibinfo {pages} {096402} (\bibinfo {year} {2017})}\BibitemShut {NoStop}%
\bibitem [{\citenamefont {Bosma}\ \emph {et~al.}(2018)\citenamefont {Bosma}, \citenamefont {Lof}, \citenamefont {Gilardoni}, \citenamefont {Zwier}, \citenamefont {Hendriks}, \citenamefont {Magnusson}, \citenamefont {Ellison}, \citenamefont {G{\"a}llstr{\"o}m}, \citenamefont {Ivanov}, \citenamefont {Son} \emph {et~al.}}]{bosma2018identification}%
  \BibitemOpen
  \bibfield  {author} {\bibinfo {author} {\bibfnamefont {T.}~\bibnamefont {Bosma}}, \bibinfo {author} {\bibfnamefont {G.~J.}\ \bibnamefont {Lof}}, \bibinfo {author} {\bibfnamefont {C.~M.}\ \bibnamefont {Gilardoni}}, \bibinfo {author} {\bibfnamefont {O.~V.}\ \bibnamefont {Zwier}}, \bibinfo {author} {\bibfnamefont {F.}~\bibnamefont {Hendriks}}, \bibinfo {author} {\bibfnamefont {B.}~\bibnamefont {Magnusson}}, \bibinfo {author} {\bibfnamefont {A.}~\bibnamefont {Ellison}}, \bibinfo {author} {\bibfnamefont {A.}~\bibnamefont {G{\"a}llstr{\"o}m}}, \bibinfo {author} {\bibfnamefont {I.~G.}\ \bibnamefont {Ivanov}}, \bibinfo {author} {\bibfnamefont {N.}~\bibnamefont {Son}}, \emph {et~al.},\ }\href@noop {} {\bibfield  {journal} {\bibinfo  {journal} {npj Quantum Information}\ }\textbf {\bibinfo {volume} {4}},\ \bibinfo {pages} {48} (\bibinfo {year} {2018})}\BibitemShut {NoStop}%
\bibitem [{\citenamefont {Bergeron}\ \emph {et~al.}(2020)\citenamefont {Bergeron}, \citenamefont {Chartrand}, \citenamefont {Kurkjian}, \citenamefont {Morse}, \citenamefont {Riemann}, \citenamefont {Abrosimov}, \citenamefont {Becker}, \citenamefont {Pohl}, \citenamefont {Thewalt},\ and\ \citenamefont {Simmons}}]{bergeron2020silicon}%
  \BibitemOpen
  \bibfield  {author} {\bibinfo {author} {\bibfnamefont {L.}~\bibnamefont {Bergeron}}, \bibinfo {author} {\bibfnamefont {C.}~\bibnamefont {Chartrand}}, \bibinfo {author} {\bibfnamefont {A.}~\bibnamefont {Kurkjian}}, \bibinfo {author} {\bibfnamefont {K.}~\bibnamefont {Morse}}, \bibinfo {author} {\bibfnamefont {H.}~\bibnamefont {Riemann}}, \bibinfo {author} {\bibfnamefont {N.}~\bibnamefont {Abrosimov}}, \bibinfo {author} {\bibfnamefont {P.}~\bibnamefont {Becker}}, \bibinfo {author} {\bibfnamefont {H.-J.}\ \bibnamefont {Pohl}}, \bibinfo {author} {\bibfnamefont {M.}~\bibnamefont {Thewalt}},\ and\ \bibinfo {author} {\bibfnamefont {S.}~\bibnamefont {Simmons}},\ }\href@noop {} {\bibfield  {journal} {\bibinfo  {journal} {PRX Quantum}\ }\textbf {\bibinfo {volume} {1}},\ \bibinfo {pages} {020301} (\bibinfo {year} {2020})}\BibitemShut {NoStop}%
\bibitem [{\citenamefont {Mitchel}\ \emph {et~al.}(2007)\citenamefont {Mitchel}, \citenamefont {Mitchell}, \citenamefont {Landis}, \citenamefont {Smith}, \citenamefont {Lee},\ and\ \citenamefont {Zvanut}}]{mitchel2007vanadium}%
  \BibitemOpen
  \bibfield  {author} {\bibinfo {author} {\bibfnamefont {W.}~\bibnamefont {Mitchel}}, \bibinfo {author} {\bibfnamefont {W.}~\bibnamefont {Mitchell}}, \bibinfo {author} {\bibfnamefont {G.}~\bibnamefont {Landis}}, \bibinfo {author} {\bibfnamefont {H.}~\bibnamefont {Smith}}, \bibinfo {author} {\bibfnamefont {W.}~\bibnamefont {Lee}},\ and\ \bibinfo {author} {\bibfnamefont {M.}~\bibnamefont {Zvanut}},\ }\href@noop {} {\bibfield  {journal} {\bibinfo  {journal} {Journal of applied physics}\ }\textbf {\bibinfo {volume} {101}} (\bibinfo {year} {2007})}\BibitemShut {NoStop}%
\bibitem [{\citenamefont {Von~Bardeleben}\ \emph {et~al.}(2019)\citenamefont {Von~Bardeleben}, \citenamefont {Zargaleh}, \citenamefont {Cantin}, \citenamefont {Gao}, \citenamefont {Biktagirov},\ and\ \citenamefont {Gerstmann}}]{von2019transition}%
  \BibitemOpen
  \bibfield  {author} {\bibinfo {author} {\bibfnamefont {H.}~\bibnamefont {Von~Bardeleben}}, \bibinfo {author} {\bibfnamefont {S.~A.}\ \bibnamefont {Zargaleh}}, \bibinfo {author} {\bibfnamefont {J.-L.}\ \bibnamefont {Cantin}}, \bibinfo {author} {\bibfnamefont {W.}~\bibnamefont {Gao}}, \bibinfo {author} {\bibfnamefont {T.}~\bibnamefont {Biktagirov}},\ and\ \bibinfo {author} {\bibfnamefont {U.}~\bibnamefont {Gerstmann}},\ }\href@noop {} {\bibfield  {journal} {\bibinfo  {journal} {Physical Review Materials}\ }\textbf {\bibinfo {volume} {3}},\ \bibinfo {pages} {124605} (\bibinfo {year} {2019})}\BibitemShut {NoStop}%
\bibitem [{\citenamefont {Wolfowicz}\ \emph {et~al.}(2017)\citenamefont {Wolfowicz}, \citenamefont {Anderson}, \citenamefont {Yeats}, \citenamefont {Whiteley}, \citenamefont {Niklas}, \citenamefont {Poluektov}, \citenamefont {Heremans},\ and\ \citenamefont {Awschalom}}]{wolfowicz2017optical}%
  \BibitemOpen
  \bibfield  {author} {\bibinfo {author} {\bibfnamefont {G.}~\bibnamefont {Wolfowicz}}, \bibinfo {author} {\bibfnamefont {C.~P.}\ \bibnamefont {Anderson}}, \bibinfo {author} {\bibfnamefont {A.~L.}\ \bibnamefont {Yeats}}, \bibinfo {author} {\bibfnamefont {S.~J.}\ \bibnamefont {Whiteley}}, \bibinfo {author} {\bibfnamefont {J.}~\bibnamefont {Niklas}}, \bibinfo {author} {\bibfnamefont {O.~G.}\ \bibnamefont {Poluektov}}, \bibinfo {author} {\bibfnamefont {F.~J.}\ \bibnamefont {Heremans}},\ and\ \bibinfo {author} {\bibfnamefont {D.~D.}\ \bibnamefont {Awschalom}},\ }\href@noop {} {\bibfield  {journal} {\bibinfo  {journal} {Nature Communications}\ }\textbf {\bibinfo {volume} {8}},\ \bibinfo {pages} {1876} (\bibinfo {year} {2017})}\BibitemShut {NoStop}%
\bibitem [{\citenamefont {Magnusson}\ \emph {et~al.}(2018)\citenamefont {Magnusson}, \citenamefont {Son}, \citenamefont {Cs{\'o}r{\'e}}, \citenamefont {G{\"a}llstr{\"o}m}, \citenamefont {Ohshima}, \citenamefont {Gali},\ and\ \citenamefont {Ivanov}}]{magnusson2018excitation}%
  \BibitemOpen
  \bibfield  {author} {\bibinfo {author} {\bibfnamefont {B.}~\bibnamefont {Magnusson}}, \bibinfo {author} {\bibfnamefont {N.~T.}\ \bibnamefont {Son}}, \bibinfo {author} {\bibfnamefont {A.}~\bibnamefont {Cs{\'o}r{\'e}}}, \bibinfo {author} {\bibfnamefont {A.}~\bibnamefont {G{\"a}llstr{\"o}m}}, \bibinfo {author} {\bibfnamefont {T.}~\bibnamefont {Ohshima}}, \bibinfo {author} {\bibfnamefont {A.}~\bibnamefont {Gali}},\ and\ \bibinfo {author} {\bibfnamefont {I.~G.}\ \bibnamefont {Ivanov}},\ }\href@noop {} {\bibfield  {journal} {\bibinfo  {journal} {Physical Review B}\ }\textbf {\bibinfo {volume} {98}},\ \bibinfo {pages} {195202} (\bibinfo {year} {2018})}\BibitemShut {NoStop}%
\bibitem [{\citenamefont {Anderson}\ \emph {et~al.}(2022)\citenamefont {Anderson}, \citenamefont {Glen}, \citenamefont {Zeledon}, \citenamefont {Bourassa}, \citenamefont {Jin}, \citenamefont {Zhu}, \citenamefont {Vorwerk}, \citenamefont {Crook}, \citenamefont {Abe}, \citenamefont {Ul-Hassan} \emph {et~al.}}]{anderson2022five}%
  \BibitemOpen
  \bibfield  {author} {\bibinfo {author} {\bibfnamefont {C.~P.}\ \bibnamefont {Anderson}}, \bibinfo {author} {\bibfnamefont {E.~O.}\ \bibnamefont {Glen}}, \bibinfo {author} {\bibfnamefont {C.}~\bibnamefont {Zeledon}}, \bibinfo {author} {\bibfnamefont {A.}~\bibnamefont {Bourassa}}, \bibinfo {author} {\bibfnamefont {Y.}~\bibnamefont {Jin}}, \bibinfo {author} {\bibfnamefont {Y.}~\bibnamefont {Zhu}}, \bibinfo {author} {\bibfnamefont {C.}~\bibnamefont {Vorwerk}}, \bibinfo {author} {\bibfnamefont {A.~L.}\ \bibnamefont {Crook}}, \bibinfo {author} {\bibfnamefont {H.}~\bibnamefont {Abe}}, \bibinfo {author} {\bibfnamefont {J.}~\bibnamefont {Ul-Hassan}}, \emph {et~al.},\ }\href@noop {} {\bibfield  {journal} {\bibinfo  {journal} {Science Advances}\ }\textbf {\bibinfo {volume} {8}},\ \bibinfo {pages} {eabm5912} (\bibinfo {year} {2022})}\BibitemShut {NoStop}%
\bibitem [{\citenamefont {Tissot}\ and\ \citenamefont {Burkard}(2021{\natexlab{b}})}]{tissot2021hyperfine}%
  \BibitemOpen
  \bibfield  {author} {\bibinfo {author} {\bibfnamefont {B.}~\bibnamefont {Tissot}}\ and\ \bibinfo {author} {\bibfnamefont {G.}~\bibnamefont {Burkard}},\ }\href@noop {} {\bibfield  {journal} {\bibinfo  {journal} {Physical Review B}\ }\textbf {\bibinfo {volume} {104}},\ \bibinfo {pages} {064102} (\bibinfo {year} {2021}{\natexlab{b}})}\BibitemShut {NoStop}%
\bibitem [{\citenamefont {Gilardoni}\ \emph {et~al.}(2021)\citenamefont {Gilardoni}, \citenamefont {Ion}, \citenamefont {Hendriks}, \citenamefont {Trupke},\ and\ \citenamefont {van~der Wal}}]{gilardoni2021hyperfine}%
  \BibitemOpen
  \bibfield  {author} {\bibinfo {author} {\bibfnamefont {C.~M.}\ \bibnamefont {Gilardoni}}, \bibinfo {author} {\bibfnamefont {I.}~\bibnamefont {Ion}}, \bibinfo {author} {\bibfnamefont {F.}~\bibnamefont {Hendriks}}, \bibinfo {author} {\bibfnamefont {M.}~\bibnamefont {Trupke}},\ and\ \bibinfo {author} {\bibfnamefont {C.~H.}\ \bibnamefont {van~der Wal}},\ }\href@noop {} {\bibfield  {journal} {\bibinfo  {journal} {New Journal of Physics}\ }\textbf {\bibinfo {volume} {23}},\ \bibinfo {pages} {083010} (\bibinfo {year} {2021})}\BibitemShut {NoStop}%
\bibitem [{\citenamefont {Bayliss}\ \emph {et~al.}(2020)\citenamefont {Bayliss}, \citenamefont {Laorenza}, \citenamefont {Mintun}, \citenamefont {Kovos}, \citenamefont {Freedman},\ and\ \citenamefont {Awschalom}}]{bayliss2020optically}%
  \BibitemOpen
  \bibfield  {author} {\bibinfo {author} {\bibfnamefont {S.}~\bibnamefont {Bayliss}}, \bibinfo {author} {\bibfnamefont {D.}~\bibnamefont {Laorenza}}, \bibinfo {author} {\bibfnamefont {P.}~\bibnamefont {Mintun}}, \bibinfo {author} {\bibfnamefont {B.}~\bibnamefont {Kovos}}, \bibinfo {author} {\bibfnamefont {D.}~\bibnamefont {Freedman}},\ and\ \bibinfo {author} {\bibfnamefont {D.}~\bibnamefont {Awschalom}},\ }\href@noop {} {\bibfield  {journal} {\bibinfo  {journal} {Science}\ }\textbf {\bibinfo {volume} {370}},\ \bibinfo {pages} {1309} (\bibinfo {year} {2020})}\BibitemShut {NoStop}%
\bibitem [{\citenamefont {Diler}\ \emph {et~al.}(2020)\citenamefont {Diler}, \citenamefont {Whiteley}, \citenamefont {Anderson}, \citenamefont {Wolfowicz}, \citenamefont {Wesson}, \citenamefont {Bielejec}, \citenamefont {Joseph~Heremans},\ and\ \citenamefont {Awschalom}}]{diler2020coherent}%
  \BibitemOpen
  \bibfield  {author} {\bibinfo {author} {\bibfnamefont {B.}~\bibnamefont {Diler}}, \bibinfo {author} {\bibfnamefont {S.~J.}\ \bibnamefont {Whiteley}}, \bibinfo {author} {\bibfnamefont {C.~P.}\ \bibnamefont {Anderson}}, \bibinfo {author} {\bibfnamefont {G.}~\bibnamefont {Wolfowicz}}, \bibinfo {author} {\bibfnamefont {M.~E.}\ \bibnamefont {Wesson}}, \bibinfo {author} {\bibfnamefont {E.~S.}\ \bibnamefont {Bielejec}}, \bibinfo {author} {\bibfnamefont {F.}~\bibnamefont {Joseph~Heremans}},\ and\ \bibinfo {author} {\bibfnamefont {D.~D.}\ \bibnamefont {Awschalom}},\ }\href@noop {} {\bibfield  {journal} {\bibinfo  {journal} {NPJ quantum information}\ }\textbf {\bibinfo {volume} {6}},\ \bibinfo {pages} {11} (\bibinfo {year} {2020})}\BibitemShut {NoStop}%
\bibitem [{\citenamefont {Gilardoni}\ \emph {et~al.}(2020)\citenamefont {Gilardoni}, \citenamefont {Bosma}, \citenamefont {Van~Hien}, \citenamefont {Hendriks}, \citenamefont {Magnusson}, \citenamefont {Ellison}, \citenamefont {Ivanov}, \citenamefont {Son},\ and\ \citenamefont {van~der Wal}}]{gilardoni2020spin}%
  \BibitemOpen
  \bibfield  {author} {\bibinfo {author} {\bibfnamefont {C.~M.}\ \bibnamefont {Gilardoni}}, \bibinfo {author} {\bibfnamefont {T.}~\bibnamefont {Bosma}}, \bibinfo {author} {\bibfnamefont {D.}~\bibnamefont {Van~Hien}}, \bibinfo {author} {\bibfnamefont {F.}~\bibnamefont {Hendriks}}, \bibinfo {author} {\bibfnamefont {B.}~\bibnamefont {Magnusson}}, \bibinfo {author} {\bibfnamefont {A.}~\bibnamefont {Ellison}}, \bibinfo {author} {\bibfnamefont {I.~G.}\ \bibnamefont {Ivanov}}, \bibinfo {author} {\bibfnamefont {N.}~\bibnamefont {Son}},\ and\ \bibinfo {author} {\bibfnamefont {C.~H.}\ \bibnamefont {van~der Wal}},\ }\href@noop {} {\bibfield  {journal} {\bibinfo  {journal} {New Journal of Physics}\ }\textbf {\bibinfo {volume} {22}},\ \bibinfo {pages} {103051} (\bibinfo {year} {2020})}\BibitemShut {NoStop}%
\bibitem [{\citenamefont {Shrivastava}(1983)}]{shrivastava1983theory}%
  \BibitemOpen
  \bibfield  {author} {\bibinfo {author} {\bibfnamefont {K.}~\bibnamefont {Shrivastava}},\ }\href@noop {} {\bibfield  {journal} {\bibinfo  {journal} {physica status solidi (b)}\ }\textbf {\bibinfo {volume} {117}},\ \bibinfo {pages} {437} (\bibinfo {year} {1983})}\BibitemShut {NoStop}%
\bibitem [{\citenamefont {Abragam}\ and\ \citenamefont {Bleaney}(1970)}]{abragam1970electron}%
  \BibitemOpen
  \bibfield  {author} {\bibinfo {author} {\bibfnamefont {A.}~\bibnamefont {Abragam}}\ and\ \bibinfo {author} {\bibfnamefont {B.}~\bibnamefont {Bleaney}},\ }\href@noop {} {\emph {\bibinfo {title} {Electron paramagnetic resonance of transition ions}}}\ (\bibinfo  {publisher} {Clarendon P.},\ \bibinfo {year} {1970})\BibitemShut {NoStop}%
\bibitem [{\citenamefont {Orbach}(1961)}]{orbach1961spin}%
  \BibitemOpen
  \bibfield  {author} {\bibinfo {author} {\bibfnamefont {R.}~\bibnamefont {Orbach}},\ }\href@noop {} {\bibfield  {journal} {\bibinfo  {journal} {Proceedings of the Royal Society of London. Series A. Mathematical and Physical Sciences}\ }\textbf {\bibinfo {volume} {264}},\ \bibinfo {pages} {458} (\bibinfo {year} {1961})}\BibitemShut {NoStop}%
\bibitem [{\citenamefont {Jahnke}\ \emph {et~al.}(2015)\citenamefont {Jahnke}, \citenamefont {Sipahigil}, \citenamefont {Binder}, \citenamefont {Doherty}, \citenamefont {Metsch}, \citenamefont {Rogers}, \citenamefont {Manson}, \citenamefont {Lukin},\ and\ \citenamefont {Jelezko}}]{jahnke2015electron}%
  \BibitemOpen
  \bibfield  {author} {\bibinfo {author} {\bibfnamefont {K.~D.}\ \bibnamefont {Jahnke}}, \bibinfo {author} {\bibfnamefont {A.}~\bibnamefont {Sipahigil}}, \bibinfo {author} {\bibfnamefont {J.~M.}\ \bibnamefont {Binder}}, \bibinfo {author} {\bibfnamefont {M.~W.}\ \bibnamefont {Doherty}}, \bibinfo {author} {\bibfnamefont {M.}~\bibnamefont {Metsch}}, \bibinfo {author} {\bibfnamefont {L.~J.}\ \bibnamefont {Rogers}}, \bibinfo {author} {\bibfnamefont {N.~B.}\ \bibnamefont {Manson}}, \bibinfo {author} {\bibfnamefont {M.~D.}\ \bibnamefont {Lukin}},\ and\ \bibinfo {author} {\bibfnamefont {F.}~\bibnamefont {Jelezko}},\ }\href@noop {} {\bibfield  {journal} {\bibinfo  {journal} {New Journal of Physics}\ }\textbf {\bibinfo {volume} {17}},\ \bibinfo {pages} {043011} (\bibinfo {year} {2015})}\BibitemShut {NoStop}%
\bibitem [{\citenamefont {Pingault}\ \emph {et~al.}(2017)\citenamefont {Pingault}, \citenamefont {Jarausch}, \citenamefont {Hepp}, \citenamefont {Klintberg}, \citenamefont {Becker}, \citenamefont {Markham}, \citenamefont {Becher},\ and\ \citenamefont {Atat{\"u}re}}]{pingault2017coherent}%
  \BibitemOpen
  \bibfield  {author} {\bibinfo {author} {\bibfnamefont {B.}~\bibnamefont {Pingault}}, \bibinfo {author} {\bibfnamefont {D.-D.}\ \bibnamefont {Jarausch}}, \bibinfo {author} {\bibfnamefont {C.}~\bibnamefont {Hepp}}, \bibinfo {author} {\bibfnamefont {L.}~\bibnamefont {Klintberg}}, \bibinfo {author} {\bibfnamefont {J.~N.}\ \bibnamefont {Becker}}, \bibinfo {author} {\bibfnamefont {M.}~\bibnamefont {Markham}}, \bibinfo {author} {\bibfnamefont {C.}~\bibnamefont {Becher}},\ and\ \bibinfo {author} {\bibfnamefont {M.}~\bibnamefont {Atat{\"u}re}},\ }\href@noop {} {\bibfield  {journal} {\bibinfo  {journal} {Nature Communications}\ }\textbf {\bibinfo {volume} {8}},\ \bibinfo {pages} {15579} (\bibinfo {year} {2017})}\BibitemShut {NoStop}%
\bibitem [{\citenamefont {Trusheim}\ \emph {et~al.}(2020)\citenamefont {Trusheim}, \citenamefont {Pingault}, \citenamefont {Wan}, \citenamefont {G{\"u}ndo{\u{g}}an}, \citenamefont {De~Santis}, \citenamefont {Debroux}, \citenamefont {Gangloff}, \citenamefont {Purser}, \citenamefont {Chen}, \citenamefont {Walsh} \emph {et~al.}}]{trusheim2020transform}%
  \BibitemOpen
  \bibfield  {author} {\bibinfo {author} {\bibfnamefont {M.~E.}\ \bibnamefont {Trusheim}}, \bibinfo {author} {\bibfnamefont {B.}~\bibnamefont {Pingault}}, \bibinfo {author} {\bibfnamefont {N.~H.}\ \bibnamefont {Wan}}, \bibinfo {author} {\bibfnamefont {M.}~\bibnamefont {G{\"u}ndo{\u{g}}an}}, \bibinfo {author} {\bibfnamefont {L.}~\bibnamefont {De~Santis}}, \bibinfo {author} {\bibfnamefont {R.}~\bibnamefont {Debroux}}, \bibinfo {author} {\bibfnamefont {D.}~\bibnamefont {Gangloff}}, \bibinfo {author} {\bibfnamefont {C.}~\bibnamefont {Purser}}, \bibinfo {author} {\bibfnamefont {K.~C.}\ \bibnamefont {Chen}}, \bibinfo {author} {\bibfnamefont {M.}~\bibnamefont {Walsh}}, \emph {et~al.},\ }\href@noop {} {\bibfield  {journal} {\bibinfo  {journal} {Physical Review Letters}\ }\textbf {\bibinfo {volume} {124}},\ \bibinfo {pages} {023602} (\bibinfo {year} {2020})}\BibitemShut {NoStop}%
\bibitem [{\citenamefont {Kuruma}\ \emph {et~al.}(2023)\citenamefont {Kuruma}, \citenamefont {Pingault}, \citenamefont {Chia}, \citenamefont {Haas}, \citenamefont {Joe}, \citenamefont {Assumpcao}, \citenamefont {Ding}, \citenamefont {Jin}, \citenamefont {Xin}, \citenamefont {Yeh} \emph {et~al.}}]{kuruma2023engineering}%
  \BibitemOpen
  \bibfield  {author} {\bibinfo {author} {\bibfnamefont {K.}~\bibnamefont {Kuruma}}, \bibinfo {author} {\bibfnamefont {B.}~\bibnamefont {Pingault}}, \bibinfo {author} {\bibfnamefont {C.}~\bibnamefont {Chia}}, \bibinfo {author} {\bibfnamefont {M.}~\bibnamefont {Haas}}, \bibinfo {author} {\bibfnamefont {G.~D.}\ \bibnamefont {Joe}}, \bibinfo {author} {\bibfnamefont {D.~R.}\ \bibnamefont {Assumpcao}}, \bibinfo {author} {\bibfnamefont {S.~W.}\ \bibnamefont {Ding}}, \bibinfo {author} {\bibfnamefont {C.}~\bibnamefont {Jin}}, \bibinfo {author} {\bibfnamefont {C.}~\bibnamefont {Xin}}, \bibinfo {author} {\bibfnamefont {M.}~\bibnamefont {Yeh}}, \emph {et~al.},\ }\href@noop {} {\bibfield  {journal} {\bibinfo  {journal} {arXiv preprint arXiv:2310.06236}\ } (\bibinfo {year} {2023})}\BibitemShut {NoStop}%
\bibitem [{\citenamefont {Klotz}\ \emph {et~al.}(2022)\citenamefont {Klotz}, \citenamefont {Fehler}, \citenamefont {Waltrich}, \citenamefont {Steiger}, \citenamefont {H{\"a}u{\ss}ler}, \citenamefont {Reddy}, \citenamefont {Kulikova}, \citenamefont {Davydov}, \citenamefont {Agafonov}, \citenamefont {Doherty} \emph {et~al.}}]{klotz2022prolonged}%
  \BibitemOpen
  \bibfield  {author} {\bibinfo {author} {\bibfnamefont {M.}~\bibnamefont {Klotz}}, \bibinfo {author} {\bibfnamefont {K.~G.}\ \bibnamefont {Fehler}}, \bibinfo {author} {\bibfnamefont {R.}~\bibnamefont {Waltrich}}, \bibinfo {author} {\bibfnamefont {E.}~\bibnamefont {Steiger}}, \bibinfo {author} {\bibfnamefont {S.}~\bibnamefont {H{\"a}u{\ss}ler}}, \bibinfo {author} {\bibfnamefont {P.}~\bibnamefont {Reddy}}, \bibinfo {author} {\bibfnamefont {L.~F.}\ \bibnamefont {Kulikova}}, \bibinfo {author} {\bibfnamefont {V.~A.}\ \bibnamefont {Davydov}}, \bibinfo {author} {\bibfnamefont {V.~N.}\ \bibnamefont {Agafonov}}, \bibinfo {author} {\bibfnamefont {M.~W.}\ \bibnamefont {Doherty}}, \emph {et~al.},\ }\href@noop {} {\bibfield  {journal} {\bibinfo  {journal} {Physical Review Letters}\ }\textbf {\bibinfo {volume} {128}},\ \bibinfo {pages} {153602} (\bibinfo {year} {2022})}\BibitemShut {NoStop}%
\bibitem [{\citenamefont {Meesala}\ \emph {et~al.}(2018)\citenamefont {Meesala}, \citenamefont {Sohn}, \citenamefont {Pingault}, \citenamefont {Shao}, \citenamefont {Atikian}, \citenamefont {Holzgrafe}, \citenamefont {G{\"u}ndo{\u{g}}an}, \citenamefont {Stavrakas}, \citenamefont {Sipahigil}, \citenamefont {Chia} \emph {et~al.}}]{meesala2018strain}%
  \BibitemOpen
  \bibfield  {author} {\bibinfo {author} {\bibfnamefont {S.}~\bibnamefont {Meesala}}, \bibinfo {author} {\bibfnamefont {Y.-I.}\ \bibnamefont {Sohn}}, \bibinfo {author} {\bibfnamefont {B.}~\bibnamefont {Pingault}}, \bibinfo {author} {\bibfnamefont {L.}~\bibnamefont {Shao}}, \bibinfo {author} {\bibfnamefont {H.~A.}\ \bibnamefont {Atikian}}, \bibinfo {author} {\bibfnamefont {J.}~\bibnamefont {Holzgrafe}}, \bibinfo {author} {\bibfnamefont {M.}~\bibnamefont {G{\"u}ndo{\u{g}}an}}, \bibinfo {author} {\bibfnamefont {C.}~\bibnamefont {Stavrakas}}, \bibinfo {author} {\bibfnamefont {A.}~\bibnamefont {Sipahigil}}, \bibinfo {author} {\bibfnamefont {C.}~\bibnamefont {Chia}}, \emph {et~al.},\ }\href@noop {} {\bibfield  {journal} {\bibinfo  {journal} {Physical Review B}\ }\textbf {\bibinfo {volume} {97}},\ \bibinfo {pages} {205444} (\bibinfo {year} {2018})}\BibitemShut {NoStop}%
\bibitem [{\citenamefont {Sohn}\ \emph {et~al.}(2018)\citenamefont {Sohn}, \citenamefont {Meesala}, \citenamefont {Pingault}, \citenamefont {Atikian}, \citenamefont {Holzgrafe}, \citenamefont {G{\"u}ndo{\u{g}}an}, \citenamefont {Stavrakas}, \citenamefont {Stanley}, \citenamefont {Sipahigil}, \citenamefont {Choi} \emph {et~al.}}]{sohn2018controlling}%
  \BibitemOpen
  \bibfield  {author} {\bibinfo {author} {\bibfnamefont {Y.-I.}\ \bibnamefont {Sohn}}, \bibinfo {author} {\bibfnamefont {S.}~\bibnamefont {Meesala}}, \bibinfo {author} {\bibfnamefont {B.}~\bibnamefont {Pingault}}, \bibinfo {author} {\bibfnamefont {H.~A.}\ \bibnamefont {Atikian}}, \bibinfo {author} {\bibfnamefont {J.}~\bibnamefont {Holzgrafe}}, \bibinfo {author} {\bibfnamefont {M.}~\bibnamefont {G{\"u}ndo{\u{g}}an}}, \bibinfo {author} {\bibfnamefont {C.}~\bibnamefont {Stavrakas}}, \bibinfo {author} {\bibfnamefont {M.~J.}\ \bibnamefont {Stanley}}, \bibinfo {author} {\bibfnamefont {A.}~\bibnamefont {Sipahigil}}, \bibinfo {author} {\bibfnamefont {J.}~\bibnamefont {Choi}}, \emph {et~al.},\ }\href@noop {} {\bibfield  {journal} {\bibinfo  {journal} {Nature Communications}\ }\textbf {\bibinfo {volume} {9}},\ \bibinfo {pages} {2012} (\bibinfo {year} {2018})}\BibitemShut {NoStop}%
\bibitem [{\citenamefont {Guo}\ \emph {et~al.}(2023)\citenamefont {Guo}, \citenamefont {Stramma}, \citenamefont {Li}, \citenamefont {Roth}, \citenamefont {Huang}, \citenamefont {Jin}, \citenamefont {Parker}, \citenamefont {Mart{\'\i}nez}, \citenamefont {Shofer}, \citenamefont {Michaels} \emph {et~al.}}]{guo2023microwave}%
  \BibitemOpen
  \bibfield  {author} {\bibinfo {author} {\bibfnamefont {X.}~\bibnamefont {Guo}}, \bibinfo {author} {\bibfnamefont {A.~M.}\ \bibnamefont {Stramma}}, \bibinfo {author} {\bibfnamefont {Z.}~\bibnamefont {Li}}, \bibinfo {author} {\bibfnamefont {W.~G.}\ \bibnamefont {Roth}}, \bibinfo {author} {\bibfnamefont {B.}~\bibnamefont {Huang}}, \bibinfo {author} {\bibfnamefont {Y.}~\bibnamefont {Jin}}, \bibinfo {author} {\bibfnamefont {R.~A.}\ \bibnamefont {Parker}}, \bibinfo {author} {\bibfnamefont {J.~A.}\ \bibnamefont {Mart{\'\i}nez}}, \bibinfo {author} {\bibfnamefont {N.}~\bibnamefont {Shofer}}, \bibinfo {author} {\bibfnamefont {C.~P.}\ \bibnamefont {Michaels}}, \emph {et~al.},\ }\href@noop {} {\bibfield  {journal} {\bibinfo  {journal} {Physical Review X}\ }\textbf {\bibinfo {volume} {13}},\ \bibinfo {pages} {041037} (\bibinfo {year} {2023})}\BibitemShut {NoStop}%
\bibitem [{\citenamefont {Stas}\ \emph {et~al.}(2022)\citenamefont {Stas}, \citenamefont {Huan}, \citenamefont {Machielse}, \citenamefont {Knall}, \citenamefont {Suleymanzade}, \citenamefont {Pingault}, \citenamefont {Sutula}, \citenamefont {Ding}, \citenamefont {Knaut}, \citenamefont {Assumpcao} \emph {et~al.}}]{stas2022robust}%
  \BibitemOpen
  \bibfield  {author} {\bibinfo {author} {\bibfnamefont {P.-J.}\ \bibnamefont {Stas}}, \bibinfo {author} {\bibfnamefont {Y.~Q.}\ \bibnamefont {Huan}}, \bibinfo {author} {\bibfnamefont {B.}~\bibnamefont {Machielse}}, \bibinfo {author} {\bibfnamefont {E.~N.}\ \bibnamefont {Knall}}, \bibinfo {author} {\bibfnamefont {A.}~\bibnamefont {Suleymanzade}}, \bibinfo {author} {\bibfnamefont {B.}~\bibnamefont {Pingault}}, \bibinfo {author} {\bibfnamefont {M.}~\bibnamefont {Sutula}}, \bibinfo {author} {\bibfnamefont {S.~W.}\ \bibnamefont {Ding}}, \bibinfo {author} {\bibfnamefont {C.~M.}\ \bibnamefont {Knaut}}, \bibinfo {author} {\bibfnamefont {D.~R.}\ \bibnamefont {Assumpcao}}, \emph {et~al.},\ }\href@noop {} {\bibfield  {journal} {\bibinfo  {journal} {Science}\ }\textbf {\bibinfo {volume} {378}},\ \bibinfo {pages} {557} (\bibinfo {year} {2022})}\BibitemShut {NoStop}%
\bibitem [{\citenamefont {Tissot}\ \emph {et~al.}(2024)\citenamefont {Tissot}, \citenamefont {Udvarhelyi}, \citenamefont {Gali},\ and\ \citenamefont {Burkard}}]{tissot2024strain}%
  \BibitemOpen
  \bibfield  {author} {\bibinfo {author} {\bibfnamefont {B.}~\bibnamefont {Tissot}}, \bibinfo {author} {\bibfnamefont {P.}~\bibnamefont {Udvarhelyi}}, \bibinfo {author} {\bibfnamefont {A.}~\bibnamefont {Gali}},\ and\ \bibinfo {author} {\bibfnamefont {G.}~\bibnamefont {Burkard}},\ }\href@noop {} {\bibfield  {journal} {\bibinfo  {journal} {Physical Review B}\ }\textbf {\bibinfo {volume} {109}},\ \bibinfo {pages} {054111} (\bibinfo {year} {2024})}\BibitemShut {NoStop}%
\bibitem [{\citenamefont {Rosenthal}\ \emph {et~al.}(2023)\citenamefont {Rosenthal}, \citenamefont {Anderson}, \citenamefont {Kleidermacher}, \citenamefont {Stein}, \citenamefont {Lee}, \citenamefont {Grzesik}, \citenamefont {Scuri}, \citenamefont {Rugar}, \citenamefont {Riedel}, \citenamefont {Aghaeimeibodi} \emph {et~al.}}]{rosenthal2023microwave}%
  \BibitemOpen
  \bibfield  {author} {\bibinfo {author} {\bibfnamefont {E.~I.}\ \bibnamefont {Rosenthal}}, \bibinfo {author} {\bibfnamefont {C.~P.}\ \bibnamefont {Anderson}}, \bibinfo {author} {\bibfnamefont {H.~C.}\ \bibnamefont {Kleidermacher}}, \bibinfo {author} {\bibfnamefont {A.~J.}\ \bibnamefont {Stein}}, \bibinfo {author} {\bibfnamefont {H.}~\bibnamefont {Lee}}, \bibinfo {author} {\bibfnamefont {J.}~\bibnamefont {Grzesik}}, \bibinfo {author} {\bibfnamefont {G.}~\bibnamefont {Scuri}}, \bibinfo {author} {\bibfnamefont {A.~E.}\ \bibnamefont {Rugar}}, \bibinfo {author} {\bibfnamefont {D.}~\bibnamefont {Riedel}}, \bibinfo {author} {\bibfnamefont {S.}~\bibnamefont {Aghaeimeibodi}}, \emph {et~al.},\ }\href@noop {} {\bibfield  {journal} {\bibinfo  {journal} {Physical Review X}\ }\textbf {\bibinfo {volume} {13}},\ \bibinfo {pages} {031022} (\bibinfo {year} {2023})}\BibitemShut {NoStop}%
\bibitem [{\citenamefont {Onizhuk}\ \emph {et~al.}(2021)\citenamefont {Onizhuk}, \citenamefont {Miao}, \citenamefont {Blanton}, \citenamefont {Ma}, \citenamefont {Anderson}, \citenamefont {Bourassa}, \citenamefont {Awschalom},\ and\ \citenamefont {Galli}}]{onizhuk2021probing}%
  \BibitemOpen
  \bibfield  {author} {\bibinfo {author} {\bibfnamefont {M.}~\bibnamefont {Onizhuk}}, \bibinfo {author} {\bibfnamefont {K.~C.}\ \bibnamefont {Miao}}, \bibinfo {author} {\bibfnamefont {J.~P.}\ \bibnamefont {Blanton}}, \bibinfo {author} {\bibfnamefont {H.}~\bibnamefont {Ma}}, \bibinfo {author} {\bibfnamefont {C.~P.}\ \bibnamefont {Anderson}}, \bibinfo {author} {\bibfnamefont {A.}~\bibnamefont {Bourassa}}, \bibinfo {author} {\bibfnamefont {D.~D.}\ \bibnamefont {Awschalom}},\ and\ \bibinfo {author} {\bibfnamefont {G.}~\bibnamefont {Galli}},\ }\href@noop {} {\bibfield  {journal} {\bibinfo  {journal} {PRX Quantum}\ }\textbf {\bibinfo {volume} {2}},\ \bibinfo {pages} {010311} (\bibinfo {year} {2021})}\BibitemShut {NoStop}%
\bibitem [{\citenamefont {Adambukulam}\ \emph {et~al.}(2024)\citenamefont {Adambukulam}, \citenamefont {Johnson}, \citenamefont {Morello},\ and\ \citenamefont {Laucht}}]{adambukulam2024hyperfine}%
  \BibitemOpen
  \bibfield  {author} {\bibinfo {author} {\bibfnamefont {C.}~\bibnamefont {Adambukulam}}, \bibinfo {author} {\bibfnamefont {B.}~\bibnamefont {Johnson}}, \bibinfo {author} {\bibfnamefont {A.}~\bibnamefont {Morello}},\ and\ \bibinfo {author} {\bibfnamefont {A.}~\bibnamefont {Laucht}},\ }\href@noop {} {\bibfield  {journal} {\bibinfo  {journal} {Physical Review Letters}\ }\textbf {\bibinfo {volume} {132}},\ \bibinfo {pages} {060603} (\bibinfo {year} {2024})}\BibitemShut {NoStop}%
\bibitem [{\citenamefont {Parker}\ \emph {et~al.}(2023)\citenamefont {Parker}, \citenamefont {Arjona~Mart{\'\i}nez}, \citenamefont {Chen}, \citenamefont {Stramma}, \citenamefont {Harris}, \citenamefont {Michaels}, \citenamefont {Trusheim}, \citenamefont {Hayhurst~Appel}, \citenamefont {Purser}, \citenamefont {Roth} \emph {et~al.}}]{parker2023diamond}%
  \BibitemOpen
  \bibfield  {author} {\bibinfo {author} {\bibfnamefont {R.~A.}\ \bibnamefont {Parker}}, \bibinfo {author} {\bibfnamefont {J.}~\bibnamefont {Arjona~Mart{\'\i}nez}}, \bibinfo {author} {\bibfnamefont {K.~C.}\ \bibnamefont {Chen}}, \bibinfo {author} {\bibfnamefont {A.~M.}\ \bibnamefont {Stramma}}, \bibinfo {author} {\bibfnamefont {I.~B.}\ \bibnamefont {Harris}}, \bibinfo {author} {\bibfnamefont {C.~P.}\ \bibnamefont {Michaels}}, \bibinfo {author} {\bibfnamefont {M.~E.}\ \bibnamefont {Trusheim}}, \bibinfo {author} {\bibfnamefont {M.}~\bibnamefont {Hayhurst~Appel}}, \bibinfo {author} {\bibfnamefont {C.~M.}\ \bibnamefont {Purser}}, \bibinfo {author} {\bibfnamefont {W.~G.}\ \bibnamefont {Roth}}, \emph {et~al.},\ }\href@noop {} {\bibfield  {journal} {\bibinfo  {journal} {Nature Photonics}\ ,\ \bibinfo {pages} {1}} (\bibinfo {year} {2023})}\BibitemShut {NoStop}%
\bibitem [{\citenamefont {Gross}(2021)}]{gross2021designing}%
  \BibitemOpen
  \bibfield  {author} {\bibinfo {author} {\bibfnamefont {J.~A.}\ \bibnamefont {Gross}},\ }\href@noop {} {\bibfield  {journal} {\bibinfo  {journal} {Physical Review Letters}\ }\textbf {\bibinfo {volume} {127}},\ \bibinfo {pages} {010504} (\bibinfo {year} {2021})}\BibitemShut {NoStop}%
\bibitem [{\citenamefont {Michaels}\ \emph {et~al.}(2021)\citenamefont {Michaels}, \citenamefont {Mart{\'\i}nez}, \citenamefont {Debroux}, \citenamefont {Parker}, \citenamefont {Stramma}, \citenamefont {Huber}, \citenamefont {Purser}, \citenamefont {Atat{\"u}re},\ and\ \citenamefont {Gangloff}}]{michaels2021multidimensional}%
  \BibitemOpen
  \bibfield  {author} {\bibinfo {author} {\bibfnamefont {C.~P.}\ \bibnamefont {Michaels}}, \bibinfo {author} {\bibfnamefont {J.~A.}\ \bibnamefont {Mart{\'\i}nez}}, \bibinfo {author} {\bibfnamefont {R.}~\bibnamefont {Debroux}}, \bibinfo {author} {\bibfnamefont {R.~A.}\ \bibnamefont {Parker}}, \bibinfo {author} {\bibfnamefont {A.~M.}\ \bibnamefont {Stramma}}, \bibinfo {author} {\bibfnamefont {L.~I.}\ \bibnamefont {Huber}}, \bibinfo {author} {\bibfnamefont {C.~M.}\ \bibnamefont {Purser}}, \bibinfo {author} {\bibfnamefont {M.}~\bibnamefont {Atat{\"u}re}},\ and\ \bibinfo {author} {\bibfnamefont {D.~A.}\ \bibnamefont {Gangloff}},\ }\href@noop {} {\bibfield  {journal} {\bibinfo  {journal} {Quantum}\ }\textbf {\bibinfo {volume} {5}},\ \bibinfo {pages} {565} (\bibinfo {year} {2021})}\BibitemShut {NoStop}%
\bibitem [{\citenamefont {Ecker}\ \emph {et~al.}(2024)\citenamefont {Ecker}, \citenamefont {Fink}, \citenamefont {Scheidl}, \citenamefont {Sohr}, \citenamefont {Ursin}, \citenamefont {Arshad}, \citenamefont {Bonato}, \citenamefont {Cilibrizzi}, \citenamefont {Gali}, \citenamefont {Udvarhelyi} \emph {et~al.}}]{ecker2024quantum}%
  \BibitemOpen
  \bibfield  {author} {\bibinfo {author} {\bibfnamefont {S.}~\bibnamefont {Ecker}}, \bibinfo {author} {\bibfnamefont {M.}~\bibnamefont {Fink}}, \bibinfo {author} {\bibfnamefont {T.}~\bibnamefont {Scheidl}}, \bibinfo {author} {\bibfnamefont {P.}~\bibnamefont {Sohr}}, \bibinfo {author} {\bibfnamefont {R.}~\bibnamefont {Ursin}}, \bibinfo {author} {\bibfnamefont {M.~J.}\ \bibnamefont {Arshad}}, \bibinfo {author} {\bibfnamefont {C.}~\bibnamefont {Bonato}}, \bibinfo {author} {\bibfnamefont {P.}~\bibnamefont {Cilibrizzi}}, \bibinfo {author} {\bibfnamefont {A.}~\bibnamefont {Gali}}, \bibinfo {author} {\bibfnamefont {P.}~\bibnamefont {Udvarhelyi}}, \emph {et~al.},\ }\href@noop {} {\bibfield  {journal} {\bibinfo  {journal} {arXiv preprint arXiv:2403.03284}\ } (\bibinfo {year} {2024})}\BibitemShut {NoStop}%
\bibitem [{\citenamefont {MacQuarrie}\ \emph {et~al.}(2021)\citenamefont {MacQuarrie}, \citenamefont {Chartrand}, \citenamefont {Higginbottom}, \citenamefont {Morse}, \citenamefont {Karasyuk}, \citenamefont {Roorda},\ and\ \citenamefont {Simmons}}]{macquarrie2021generating}%
  \BibitemOpen
  \bibfield  {author} {\bibinfo {author} {\bibfnamefont {E.}~\bibnamefont {MacQuarrie}}, \bibinfo {author} {\bibfnamefont {C.}~\bibnamefont {Chartrand}}, \bibinfo {author} {\bibfnamefont {D.}~\bibnamefont {Higginbottom}}, \bibinfo {author} {\bibfnamefont {K.}~\bibnamefont {Morse}}, \bibinfo {author} {\bibfnamefont {V.}~\bibnamefont {Karasyuk}}, \bibinfo {author} {\bibfnamefont {S.}~\bibnamefont {Roorda}},\ and\ \bibinfo {author} {\bibfnamefont {S.}~\bibnamefont {Simmons}},\ }\href@noop {} {\bibfield  {journal} {\bibinfo  {journal} {New Journal of Physics}\ }\textbf {\bibinfo {volume} {23}},\ \bibinfo {pages} {103008} (\bibinfo {year} {2021})}\BibitemShut {NoStop}%
\bibitem [{\citenamefont {Baier}\ \emph {et~al.}(2020)\citenamefont {Baier}, \citenamefont {Bradley}, \citenamefont {Middelburg}, \citenamefont {Dobrovitski}, \citenamefont {Taminiau},\ and\ \citenamefont {Hanson}}]{baier2020orbital}%
  \BibitemOpen
  \bibfield  {author} {\bibinfo {author} {\bibfnamefont {S.}~\bibnamefont {Baier}}, \bibinfo {author} {\bibfnamefont {C.}~\bibnamefont {Bradley}}, \bibinfo {author} {\bibfnamefont {T.}~\bibnamefont {Middelburg}}, \bibinfo {author} {\bibfnamefont {V.}~\bibnamefont {Dobrovitski}}, \bibinfo {author} {\bibfnamefont {T.}~\bibnamefont {Taminiau}},\ and\ \bibinfo {author} {\bibfnamefont {R.}~\bibnamefont {Hanson}},\ }\href@noop {} {\bibfield  {journal} {\bibinfo  {journal} {Physical Review Letters}\ }\textbf {\bibinfo {volume} {125}},\ \bibinfo {pages} {193601} (\bibinfo {year} {2020})}\BibitemShut {NoStop}%
\bibitem [{\citenamefont {Harris}\ \emph {et~al.}(2023)\citenamefont {Harris}, \citenamefont {Michaels}, \citenamefont {Chen}, \citenamefont {Parker}, \citenamefont {Titze}, \citenamefont {Mart{\'\i}nez}, \citenamefont {Sutula}, \citenamefont {Christen}, \citenamefont {Stramma}, \citenamefont {Roth} \emph {et~al.}}]{harris2023hyperfine}%
  \BibitemOpen
  \bibfield  {author} {\bibinfo {author} {\bibfnamefont {I.~B.}\ \bibnamefont {Harris}}, \bibinfo {author} {\bibfnamefont {C.~P.}\ \bibnamefont {Michaels}}, \bibinfo {author} {\bibfnamefont {K.~C.}\ \bibnamefont {Chen}}, \bibinfo {author} {\bibfnamefont {R.~A.}\ \bibnamefont {Parker}}, \bibinfo {author} {\bibfnamefont {M.}~\bibnamefont {Titze}}, \bibinfo {author} {\bibfnamefont {J.~A.}\ \bibnamefont {Mart{\'\i}nez}}, \bibinfo {author} {\bibfnamefont {M.}~\bibnamefont {Sutula}}, \bibinfo {author} {\bibfnamefont {I.~R.}\ \bibnamefont {Christen}}, \bibinfo {author} {\bibfnamefont {A.~M.}\ \bibnamefont {Stramma}}, \bibinfo {author} {\bibfnamefont {W.}~\bibnamefont {Roth}}, \emph {et~al.},\ }\href@noop {} {\bibfield  {journal} {\bibinfo  {journal} {PRX Quantum}\ }\textbf {\bibinfo {volume} {4}},\ \bibinfo {pages} {040301} (\bibinfo {year} {2023})}\BibitemShut {NoStop}%
\end{thebibliography}%

\end{document}